\documentclass[sigplan,10pt,authorversion,screen]{acmart}
\usepackage{tikz}
\usepackage{amsmath}

\usepackage{filecontents}

\usepackage{textcomp} %% new

\usepackage{setspace}
\usepackage{multirow}
\usepackage{makecell}
\usepackage{verbatim}
\usepackage{paralist}
\usepackage{subcaption}
\usepackage{graphicx}
\usepackage{balance}
\usepackage{rotating}
\usepackage{pifont}
\usepackage{tikz}
\usepackage{color}
\usepackage{xcolor}
\usepackage{framed}
\usepackage[para,flushleft,online]{threeparttable}
\usepackage{paralist}
\usepackage[labelfont={bf}, textfont={}]{caption}
\usepackage{listings}
\usepackage{enumitem}
\usepackage{algorithmicx}
\usepackage{algpseudocode}
\usepackage[ruled,linesnumbered]{algorithm2e} % For algorithms
\usepackage{soul}

\definecolor{mygreen}{rgb}{0,0.6,0}
\definecolor{mygray}{rgb}{0.5,0.5,0.5}
\definecolor{mymauve}{rgb}{0.58,0,0.82}
\usepackage{def}

\copyrightyear{2023} 
\acmYear{2023} 
\setcopyright{acmlicensed}\acmConference[SOSP '23]{ACM SIGOPS 29th Symposium on Operating Systems Principles}{October 23--26, 2023}{Koblenz, Germany}
\acmBooktitle{ACM SIGOPS 29th Symposium on Operating Systems Principles (SOSP '23), October 23--26, 2023, Koblenz, Germany}
\acmPrice{15.00}
\acmDOI{10.1145/3600006.3613170}
\acmISBN{979-8-4007-0229-7/23/10}

\begin{document}
%\newpage
%\setcounter{page}{1}
%\input{rlog}

\author{Benjamin Reidys}
\affiliation{%
    \country{University of Illinois Urbana-Champaign, USA}
}
\author{Yuqi Xue}
\affiliation{%
    \country{University of Illinois Urbana-Champaign, USA}
}
\author{Daixuan Li}
\affiliation{%
    \country{University of Illinois Urbana-Champaign, USA}
}
\author{Bharat Sukhwani}
\affiliation{%
    \country{IBM T. J. Watson Research Center, Yorktown Heights, USA}
}
\author{Wen-mei Hwu}
\affiliation{%
    \country{University of Illinois Urbana-Champaign, USA}
}
\author{Deming Chen}
\affiliation{%
    \country{University of Illinois Urbana-Champaign, USA}
}
\author{Sameh Asaad}
\affiliation{%
    \country{IBM T. J. Watson Research Center, Yorktown Heights, USA}
}
\author{Jian Huang}
\affiliation{%
    \country{University of Illinois Urbana-Champaign, USA}
}
\title{\pname{}: A Software-Defined Rack-Scale Storage System with Network-Storage Co-Design} 
%\titlenote{The paper will appear in the 29th ACM Symposium on Operating Systems Principles (SOSP'23).}
%\title{\vspace{-8ex}\pname{}: An Efficient Rack-Scale Storage System \\with Software-Defined Network and Storage Co-Design\vspace{-10ex}} 
%\title{\vspace{-8ex}\pname{}: Improving the End-to-End Performance for Rack-Scale Storage Systems with Software-Defined Network and Storage Co-Design\vspace{-10ex}} 

%Relavant CCS topics

\renewcommand{\shortauthors}{B. Reidys, Y. Xue, D. Li, B. Sukhwani, W. Hwu, D. Chen, S. Asaad, and J. Huang}

\begin{abstract}
Software-defined networking (SDN) and software-defined flash (SDF) have been serving as the backbone of modern data centers. 
They are managed separately to handle I/O requests.
At first glance, this is a reasonable design by following the rack-scale hierarchical design principles. 
%However, due to the lack of coordination between SDN and SDF, it is challenging to achieve  
%end-to-end performance. 
However, it suffers from suboptimal end-to-end performance, due to the lack of coordination between SDN and SDF. 

%although they share the similar control/data plane architecture. 
%For instance, the scheduling policies configured in SDN and SDF may not be compatible with 
%each other, making it challenging for applications to achieve low and constant end-to-end latency. 
%as they provide flexibility and agility for 
%platform operators to customize the hardware resource for applications. 
	
In this paper, we co-design the SDN and SDF stack by redefining the functions of their control plane 
and data plane, and splitting up them within a new architecture named \pname{}. 
%Such a new architecture enables state sharing between SDN and SDF, and facilitates global resource management. 
\pname{} decouples the storage management functions of flash-based solid-state drives (SSDs), 
and allow the SDN to track and manage the states of SSDs in a rack.  
%and integrate appropriate indications, such as garbage collection and wear leveling into the control plane of SDN. 
Therefore, we can enable the state sharing between SDN and SDF, and facilitate global storage resource management. 
\pname{} has three major components: (1) coordinated I/O scheduling, in which it 
%Second, \pname{} 
%coordinates the effort of I/O scheduling across the software-defined networking and storage system stack by 
dynamically adjusts the I/O scheduling in the storage stack with the measured and predicted network latency, such 
that it can coordinate the effort of I/O scheduling across the network and storage stack for achieving 
predictable end-to-end performance; (2) coordinated garbage collection (GC), in which it will coordinate 
the GC activities across the SSDs in a rack to minimize their impact on incoming I/O requests; (3) rack-scale 
wear leveling, in which it enables global wear leveling among SSDs in a rack by periodically swapping data, 
for achieving improved device lifetime for the entire rack. 
%Third, \pname{} enables various coordinated garbage collection schemes among the SSDs in a rack, which 
%can maximize the performance efficiency for applications having different storage characteristics. \pname{} 
%also enables global wear leveling among SSDs for achieving improved device lifetime. 
We implement \pname{} using programmable SSDs and switch. 
Our experiments demonstrate that \pname{} can reduce the tail latency of I/O requests 
by up to 5.8$\times$ over state-of-the-art rack-scale storage systems.  
% and improve the application performance by up to {1.7}$\times$, 

\end{abstract}

\maketitle
\pagenumbering{gobble}

%\newpage
%\setcounter{page}{1}
\section{Introduction}
\label{sec:intro}
The software-defined infrastructure has become the new standard for managing data centers,
as it provides flexibility and agility for platform operators to customize hardware resources
for applications~\cite{ouyang:asplos2014, sdn:acm2016, sdi}. As the backbone technology, 
software-defined networking (SDN) allows network operators to configure and manage network resources 
through programmable switches~\cite{liu:asplos2017, pat:sigcomm2014, netcache:sosp2017, netchain:nsdi2018}.
Since SDN has demonstrated its benefits, 
%Similarly, the storage stack has deployed software-defined techniques, called 
software-defined storage 
(SDS)~\cite{ouyang:asplos2014, eno:sosp2013, zhang:sigmod2014} has also been developed. 
A typical example is software-defined flash (SDF)~\cite{ouyang:asplos2014, bjorling:fast2017, huang:fast17, blockflex:osdi2022}.  
%to allow
% multi-tenant storage services to customize management of the physical storage devices to their workload
%characteristics,
%which is known as software-defined storage (SDS). 
%In particular, because the cost
%of SSDs has been dramatically decreased, and they outperform hard disk 
%drives (HDDs) by orders of magnitude~\cite{huang-fast17, flashmap:isca2015}, SSDs are becoming the 
%mainstream choice in large-scale data centers~\cite{ssdcloud, hao:sosp2017, hyperloop:sigcomm18}. 
%%In the meantime, as the tail
%%latency of the SSD can be high and its performance can be unpredictable, which threats QoS assurance required
%%in many important services,
%And naturally, software-defined flash (SDF) has been under intensive study and wide deployment~\cite{ouyang:asplos2014, bjorling:fast2017}. 
%%Like the programmable switch in SDN,
%%the SDF decouples the storage management (i.e., flash translation layer) and data access on the flash chips,
%%and allows upper-level workloads to exploit performance characteristics of flash memory.

Similar to SDN, SDF enables upper-level software to manage the low-level flash chips for improved performance and 
resource utilization~\cite{ouyang:asplos2014,lee:fast2016, huang:fast17}. Since the cost of flash chips has 
dramatically decreased while offering orders of magnitude better performance than conventional hard disk drives (HDDs), 
they are becoming the mainstream choice in large-scale data centers~\cite{ssdcloud, hao:sosp2017, hyperloop:sigcomm18}.
%Naturally, SDF has been under intensive study and wide deployment~\cite{huang:fast17, ssdcloud,
%opencompute, perseus:fast2023}.

Both SDN and SDF have their own control plane and data plane, and provide programmability for 
developers to define and implement their policies for resource management and scheduling.
However, SDN and SDF are managed separately in modern data centers.
At first glance, this is reasonable by following the rack-scale hierarchical design principles.
However, it suffers from suboptimal end-to-end performance, due to the lack of coordination between SDN and SDF.
%First,

Although both SDN and SDF can make the best effort to
achieve their quality of service, they do not share their states and lack global information for 
storage management and scheduling, 
%the scheduling policies defined in SDN and SDS may not be compatible with each other, 
making it challenging for applications to achieve predictable end-to-end performance. 
%and satisfy their service-level objective (SLO). 
%Second, the SDN and SDF do not share their states and lack global information for storage management. 
Prior studies~\cite{eno:sosp2013, virtualdc:osdi2014} have proposed various software techniques such 
as token bucket and virtual cost for enforcing performance isolation across the rack-scale storage 
stack. However, they treat the underlying SSDs as black boxes, and 
cannot capture their hardware events, such as garbage collection (GC) and I/O scheduling in the storage stack. 
Thus, it is still hard to achieve predictable performance across the entire rack. 
%Second, as the infrastructure is scaling to support an increasing number of network and storage services,
%it is important to sustain high scalability of concurrent I/O accesses. However, our preliminary studies
%have identified that the current implementation of network packet processing and storage accesses
%are still suffering from severe scalability challenges.
%%due to the lack of coordination among cloud services,
%%network, and storage servers.
%Third, as replication has become a fundamental technique to provide
%the fault tolerance and ensure data reliability,
%%strongly consistent replication can transparently mask failures to 
%%achieve high system availability, however, it comes with a high performance cost. 
%we are still suffering from the fundamental tussle between scalability and consistency.
%Finally, as memory/storage is still struggling with scaling challenges, we have boosted up the
%computational capabilities with intensive accelerator development, the gap between compute and
%data will only be enlarged if not addressed properly.

%In this paper, we rethink the software-defined network and storage stack, and 
In this paper, we propose a new 
software-defined architecture, named \pname{}, to exploit the capabilities of SDN and SDF in a coordinated 
fashion. %to address the aforementioned challenge. 
%Our key insight is that the current SDN and SDF are lacking
%coordination and this causes severe performance, scalability, and reliability challenges. 
As both SDN and SDF share a similar architecture--the control plane is responsible for managing the 
programmable devices, and the data plane is responsible for processing I/O requests--we
%naturally, we can integrate and co-design both SDN and SDF, and redefine their functions 
can integrate and co-design both SDN and SDF, and redefine their functions 
to improve the efficiency of the entire rack-scale storage system. \pname{} does not require 
new hardware changes, as both SDN and SDF today have offered the flexibility to redefine the 
functions of their data planes. 

%Specifically, 
To develop \pname{}, we first decouple the functions of the storage management (i.e., flash translation layer)
of SDF, and integrate appropriate functions such as garbage collection and wear leveling 
into the control plane of top-of-rack (ToR) switches in the SDN. 
%We keep the 
%necessary functions such as bad block management and error correction in the SDF devices, 
%since these functions are more convenient to be handled with local SSD devices. 
Such a new software-defined architecture enables state sharing between SDN and SDF, 
and facilitates the global storage resource management in a rack.
This is compatible with storage virtualization by enabling the state tracking of 
virtualized SSD instances in SDN. 
%\begin{figure}[t]
%        \centering
%                \includegraphics[width=0.95\linewidth]{Figures/storagerack.pdf}
%		\vspace{-1.5ex}
%	%\caption{An Overview of the Software-Defined Infrastructure at Rack Scale with Network-Storage Co-Design.}
%	\caption{The overview of a rack-scale storage system.}
%        \label{fig:storagerack}
%		\vspace{-3ex}
%\end{figure}
%After the function integration of SDN and SDF, \pname{} enables the state sharing of
%I/O events on the data planes of SDN and SDF. 
Therefore, we can coordinate the efforts of I/O
scheduling across the entire rack. \pname{} tracks the elapsed time in the programmable 
switches with In-band Network Telemetry (INT)~\cite{int}, adapts the I/O scheduling in the data plane of SDF, 
and predicts the response time from the storage devices back to the client. 
Therefore, \mbox{\pname{}} can manage the end-to-end latency and offer predictable performance.
% Therefore, \pname{} can manage the end-to-end delay for the I/O request and offer predictable 
% performance to end users.

%henceforth, the SDN and SDF can control the end-to-end delay
%for the request and provide precise feedback to the I/O scheduling in the next stage. 
%Such a QoS mechanism
%with global scheduling information ensures the end-to-end I/O performance and enables intelligent
%decision making in advance, such as early I/O cancellation or redirection.
%%Finally, we propose to extend the abstraction of SDN (e.g., P4 language for programmable switch) for
%%the new software-defined architecture, which enables platform operators to fulfill more advanced
%%features such as automate data replication and migration, fault tolerance, and coordinated garbage collection
%%proposed in this project.

%With the new software-defined architecture, we further rethink storage management at rack-scale. 
\pname{} further enables coordinated GC among the rack of SSDs to minimize the impact of GC on application performance. 
%With \pname{}, the control plane of the SDN schedules the coordinated GC, as it 
\pname{} has the global information of the storage states, which provides the convenience to 
coordinate GC events among all the SSDs in a rack. Upon GC events, \pname{} takes advantage of the data 
replicas in the same rack, and enables the ToR switch to redirect I/O requests to the other data replicas. 
Therefore, the expensive GC activities can be alleviated from the critical path. \pname{} employs different 
GC policies for different performance isolation guarantees of virtualized SSD instances. 
%\pname{} allows platform operators to customize their own GC policies according to their application characteristics. 
%For instance, \pname{} supports three types of GC schemes by default, including stop-the-world GC for homogeneous workloads that would 
%finish almost simultaneously, on-demand GC for heterogeneous workloads, and replication-based GC for latency-sensitive workloads, respectively. 
%We propose the in-network lock
%manager to scale the concurrent storage accesses. Also, we will exploit the hardware parallelism in multi-core
%CPU chips and develop near-cache accelerators to scale the network packet processing in switches.
%Furthermore, we rethink the existing data replication protocols and propose the in-network based fault tolerance
%to achieve both linear scalability and strong data consistency.
%We will apply the proposed techniques to representative multi-tenant storage services and demonstrate
%the benefits of the new proposed software-defined architecture.

\pname{} also enables rack-scale wear leveling to ensure a uniform lifetime of SSDs in a rack. 
As the write traffic to each SSD can be different, it will cause wear imbalance between SSDs. 
In addition, platform operators have to replace unhealthy or failed SSDs with new SSDs, making the wear imbalance 
even harder to manage. \pname{} develops a two-level wear leveling mechanism. It balances wear 
within each individual SSD in a storage server as well as across SSDs in the rack. Instead of swapping 
SSDs frequently, \pname{} periodically swaps the SSD that has incurred the maximum wear with the SSD that 
has the minimum rate of wear. 
%It also leverages the data replication mechanism available in modern data centers to minimize the negative performance 
%impact of data migration 

We implement \pname{} with a programmable Tofino switch and programmable SSDs 
(i.e., open-channel SSDs). We evaluate \pname{} with network traces collected from various data 
centers and a variety of data-intensive applications. Our experiments show that \pname{} reduces the tail latency 
of end-to-end I/O requests by up to 5.8$\times$, 
%and improves the average application performance by up to 1.7$\times$. 
%We also show that \pname{} 
and can achieve a uniform lifetime for a rack of SSDs without introducing much additional 
performance overhead. 
In summary, we make the following contributions in this paper. 

\begin{itemize}[leftmargin=*]
%	\vspace{-1ex}
	\item We propose a new software-defined rack-scale storage system by decoupling the 
		storage management functions of SDF, and co-designing them with SDN. 
		%This enables state sharing 
		%between SDN and SDF. 

	\item %We develop a new I/O scheduling mechanism by coordinating the efforts of I/O request scheduling
		We enable state sharing between SDN and SDF, and coordinate the efforts of I/O request scheduling 
		across the full rack for achieving predictable end-to-end performance. 

	\item We present a coordinated GC mechanism for a rack of SSDs, it enables SDN to redirect I/O requests 
		to data replicas to minimize the GC impact on storage performance. 
		%which enable platform operators to define their own 
		%garbage collection policies according to their application characteristics. 

	\item We develop a rack-scale wear leveling mechanism for ensuring the uniform lifetime of a rack of SSDs.
		%while having with minimal negative impact on the storage performance.   
	
	\item We show the benefits of \pname{} by developing a real system prototype with programmable 
		switch and SSDs. 

	\vspace{-2ex}
\end{itemize}

%We organize the rest of the paper as follows. We discuss the background and motivation in $\S$\ref{sec:motivation}. We present the design 
%and implementation of \pname{} in $\S$\ref{sec:design}, and evaluate it in $\S$\ref{sec:eval}. We present the  
%related work in $\S$\ref{sec:related}, and conclude the paper in $\S$\ref{sec:conclusion}. 

\section{Background and Motivation}
\label{sec:motivation}
%In this section, 
We first introduce the background of SDN and SDS, 
%respectively. We 
then the motivation for software-defined network/storage co-design. 
%We then discuss the problems with modern rack-scale storage systems. 

\vspace{-1ex}
\subsection{Software-Defined Networking}
\label{subsec:motiv_sdn}
Modern data centers
%the server machines are usually organized in racks and the top-of-rack (ToR) switching is used to 
%connect all these server.
have seen a trend that software-defined networking (SDN) has become the new standard for
network management, in which the programmable switch is the backbone technology that allows
platform operators to define their own packet formats and functions for processing network traffic
without affecting the line-rate throughput~\cite{consensus-routing, 
netcache:sosp2017}. SDN has been deployed in real data centers such as Alibaba cloud~\cite{cetus:nsdi2022, sailfish:sigcomm2021} 
and Google data centers~\cite{google-jupiter}.

\begin{figure}[t]
        \centering
                \includegraphics[width=0.9\linewidth]{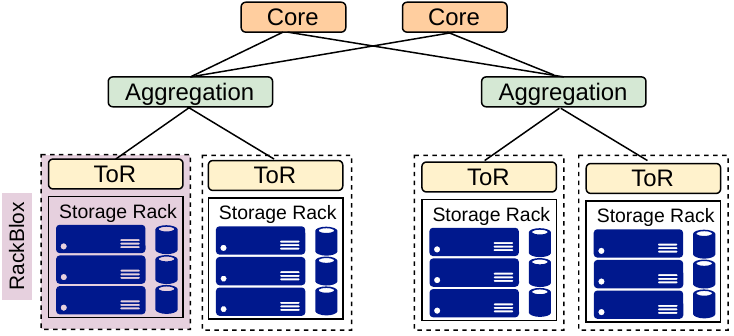}
                \vspace{-2ex}
        %\caption{An Overview of the Software-Defined Infrastructure at Rack Scale with Network-Storage Co-Design.}
        \caption{System overview of a rack-scale storage system.}
        \label{fig:storagerack}
                \vspace{-3ex}
\end{figure}

SDN has a control plane and data plane. The control plane is in charge of
network management and protocol definition, while the data plane is responsible for %flow classification and 
data transfer and run-time statistics collection. The programmable switch usually has reconfigurable hardware such as a %FPGA and
programmable ASIC that supports domain-specific languages like P4~\cite{pat:sigcomm2014}.
It supports various network flow scheduling policies for flexible traffic management and  
performance isolation~\cite{pfabric,detail,pdq}.
%\hl{This supports defining custom packet formats and custom packet processing to implement various network flow scheduling policies for flexible traffic management and  
%performance isolation~\mbox{\cite{pfabric,detail,pdq}} without affecting line-rate throughput.}
As shown in 
Figure~\ref{fig:storagerack}, all the servers in the same rack are connected by a Top-of-Rack (ToR) switch. 
These ToR switches are connected with aggregation switches and core switches in a hierarchical manner. 
%We co-design network and storage stack in a rack.  
In this paper, we focus on the ToR switch, and co-design network and storage stack in a rack. 
%We utilize this hardware trend of programmable switches and identify the
%optimization opportunities to co-design the network and storage stack at rack scale. 

\vspace{-1ex}
\subsection{Software-Defined Storage}
\label{subsec:sds}
Recent studies have shown that making software aware of the underlying storage devices 
can significantly improve the storage performance and resource efficiency~\cite{huang:fast17,wang:eurosys2014,ouyang:asplos2014}.
This is known as software-defined storage (SDS), 
%in which the storage performance~\cite{shue:eurosys2014, shue:osdi2012} can be improved
%by observing the convex-dependency between I/O requests~\cite{shue:eurosys2014, shue:osdi2012}, and predicting 
%workload patterns~\cite{ahn:hotstorage2016}. It also 
which enables data centers to unlock
the potential of storage devices by enabling the software to directly interact
with storage devices and control their internal operations.
Software-defined flash (SDF), which is built on SSDs, is a typical
example of SDS, and has seen deployment in industry data centers\mbox{~\cite{AlibabaSDF, perseus:fast2023}}. 

\begin{figure}[t]
        \centering
                \includegraphics[width=0.9\linewidth]{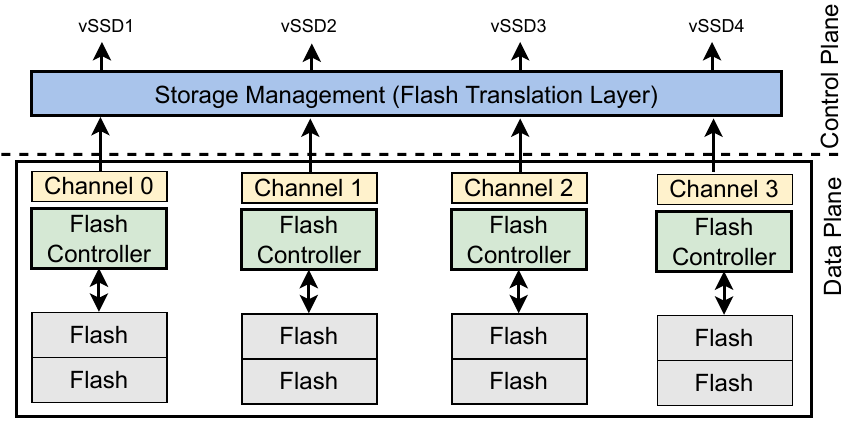}
		\vspace{-2.5ex}
        \caption{System architecture of software-defined flash.}
        \label{fig:sdf}
		\vspace{-3ex}
\end{figure}

In this paper, we focus on SDF, because flash-based SSDs are becoming indispensable parts of modern computer systems. 
An SSD has three major components: a set of flash memory packages, % that form the core of the storage, 
an SSD controller having an embedded processor with device memory, and flash controllers.
%The flash packages are organized hierarchically. 
As shown in Figure~\ref{fig:sdf}, 
each SSD has multiple channels and each channel can receive and process I/O commands independently. Each channel
is shared by multiple flash memory packages. Each package is made of multiple chips. 
%Within each chip there are multiple planes. The plane is divided into blocks, 
Each chip has multiple flash blocks. %each consisting of multiple flash pages. 
With SDF, an SSD can be virtualized into multiple virtual SSD instances (vSSDs), and each can be mapped to 
a set of SSDs, flash channels, or flash chips. 

Due to the nature of flash memory, 
%flash chips can read and write only at a page granularity, and writes can only occur to free pages.
when a free page is written, that page is no longer available for future writes until that
page is erased. However, erase operations can be performed only at block %(which has 64 or more pages) 
granularity, which are time-consuming.
Thus, writes are issued to free pages erased in advance (i.e., out-of-place write) rather than waiting
for the expensive erase operation for every write. And garbage collection (GC) will be performed later to erase 
the stale data on SSDs. Since an SSD channel cannot issue new I/O requests during GC, minimizing the 
negative impact of GC events is critical to storage performance. 
In addition, as each flash block has 
limited endurance, it is important for the blocks to age uniformly (i.e., wear leveling). 
%SSDs employ both out-of-place update and GC to overcome 
%the shortcomings of SSDs, and maintain indirections for indexing the logical-to-physical address 
%mapping in the Flash Translation Layer (FTL). 

SSDs have the Flash Translation Layer (FTL) to manage flash blocks and maintain the logical-to-physical address 
mappings. Unlike conventional SSDs that implement the FTL  
in the device firmware, SDF exposes the FTL to the upper-level software, and enables the software to manage 
the flash chips (see Figure~\ref{fig:sdf}).

%\subsection{The Gap Between SDF and SDN}
\vspace{-0.9ex}
\subsection{Why Network-Storage Co-Design}
\label{subsec:why}
In modern data centers, the SDN and SDF are managed separately to handle I/O requests across the 
network and storage stack, respectively. Such an architecture suffers 
from suboptimal performance and misses the opportunities offered by the software-defined rack for 
three major reasons. 
%This is for three reasons described as follows. 

%First, as SDN and SDF are deployed as independent components, their incoordination makes it hard to 
%achieve predictable end-to-end performance. %and fairness for applications and storage services.
First, as SDN and SDF are deployed as independent components, achieving predictable end-to-end performance is challenging, due to the lack of coordination between 
the two components.
SDN and SDF have redundant control plane policies, such as I/O scheduling, which may contradict between the network and storage stack and
break service-level objectives (SLOs). And optimizing such policies without coordination is suboptimal due to incomplete
knowledge and redundant effort. 
%%To make matters worse, the QoS policies defined in SDN and SDF could conflict with each other and easily 
%%break the service-level objective (SLO) for end users. 
%%Although the SDN and SDF provide us with the 
%%opportunities to obtain the knowledge of the low-level hardware events,
%As the network stack is also serving network traffic for other application services, such as distributed 
%computing, the I/O requests to the storage stack could be interferenced significantly. 
%%Similiarly, these I/O 
%%requests could be affected by the co-located storage services~\cite{huang:fast17,ahn:hotstorage2016}.
Ideally, as we forward I/O requests in SDN,
with the knowledge of the storage status (e.g., busy, idle, or predicted performance), it can make smarter decisions (e.g., early redirection to data replicas). Similarly, as SDF
schedules the received I/O requests, the measured network latency of these I/O requests can help
%the SDF to adjust the I/O scheduling accordingly to meet the service-level objective (SLO) for end users. 
the SDF to adjust the I/O scheduling to meet the SLO for end users.

Second, although prior studies such as IOFlow~\cite{eno:sosp2013} and VDC~\cite{virtualdc:osdi2014} proposed 
software-based methods like token bucket rate limiting to enforce the performance isolation between I/O flows, 
they cannot capture the underlying hardware events such as GC and I/O scheduling in SSDs, 
due to the lack of state sharing between SDN and SDF. And software-based coordination incurs extra network 
round-trip delay and host software overhead (see our evaluation in $\S$\ref{sec:eval}).  

Third, it is feasible to co-design and coordinate the network and storage stack today, as both programmable 
switches and programmable SSDs have enabled developers to program and configure the network and storage 
stack respectively. %It offers opportunities for network/storage co-design. 
%\hl{Furthermore, the programmable switch is optimally situated to enable rack-level management at
%line-rate. As shown in prior work, alternate solutions relying on the servers cannot support the
%same throughput as the switch and would harm end-to-end
%performance~\mbox{\cite{tea:sigcomm2020,beacoup:sigcomm2020,silkroad:sigcomm2017}.} We will further
%elaborate on the performance benefits of using a programmable switch in $\S$\mbox{\ref{sec:eval}}}

%TODO CUT Candidate
In this work, we integrate the storage management of SDF into SDN as shown in Figure~\ref{fig:overview}, while
preserving the programmability and simplicity for the new infrastructure.

\vspace{-0.5ex}
\section{\pname{} Design and Implementation}
\label{sec:design}
\label{subsec:goals}
%\pname{} provides a holistic approach that can not only achieve predictable end-to-end
%performance, but also improve the storage management at scale. %while ensuring data reliability in software-defined data centers.
%As we develop the new software-defined architecture based on modern SDN and SDF, 
%we have to overcome the following challenges. 
\pname{} provides a holistic approach that can achieve predictable end-to-end
performance and improve storage management at scale. %while ensuring data reliability in software-defined data centers.
As we develop \pname{} based on modern SDN and SDF, 
we have to overcome the following challenges. 

\vspace{-1ex}
\subsection{Design Challenges of \pname{}}

\vspace{-0.5ex}
\begin{itemize}[leftmargin=*]
\item
        	%As we co-design the network and storage stack, 
		It is unclear how the functions of
                storage management should be decoupled and placed across SDN and SDF.
                %Take the SDF as a case study in this project,
                The control plane of SDF has many functions, %of flash-based storage management,
                including wear leveling, GC, block allocation, and block management.  
                %it is not clear what is the best
                %strategy to decouple them and decide which functions should be placed at where across the SDN and SDF stacks.
                %In SDF, the flash translation layer (FTL), which was designed specifically for managing flash chips,
                %is exposed to upper-level software systems. It works as the control plane that allows software to define
                %its own storage stack. However, the FTL has many critical functions such as address translation, block
                %allocation, wear leveling, and garbage collection.
                %A simple approach is to 
		Placing all the storage functions into SDN will inevitably increase the burden of SDN. 
                Thus, we have to carefully decide the partitioning and placement of SDF functions. 
		%how to split the FTL
                %functions and where the selected FTL functions will be placed in the SDN, 
		%which could enable us to achieve
                %the predictable end-to-end performance.
                %the function partitioning between the control/data planes of SDN and SDF.

\item
		%As we exploit the software-defined devices to develop \pname{}, 
		%address the performance, scalability, and reliability issues,
                The hardware resources of programmable devices are limited.
		Specifically, the on-chip memory (tens of MBs) and compute resource are limited in programmable switches 
		and SSD controllers, due to the hardware cost and power budget. Thus, we have to carefully 
    		%define the data structures for rack-scale storage management. 
            %(BEN) a bit more targeted, since we have to maintain information in both directions.
            define the data structures for the network/storage co-design. 
                %In programmable switches or storage devices, the hardware resources such as on-chip memory
                %and processors are limited due to the cost and power constraints.
               % For instance, modern commodity switches have only tens of MBs on-chip memory. 
               % which cannot store all the metadata (hundreds of GBs) for a rack of storage devices.
                %Therefore, %as we rethink the distributed protocols with network/storage co-design,
                %we have to carefully analyze the trade-offs in different design options across the network
                %and storage stack.

\item  
		As we enable the coordinated storage management between SDN and SDF, 
		we must preserve their programmability, ease-of-use, and original advantages. 
		Thus, \pname{} should be compatible with hardware upgrades. 
		%extended to  new demands from applications and hardware upgrades. 

%\item
%        	With the new software-defined architecture, the research community lacks reference
%                application cases to fully demonstrate the its benefits.
%                Therefore, we are motivated to develop representative prototype systems and case studies to evaluate the
%                proposed research. The results and lessons learned from these studies will provide guidance to the adoption
%                of the new software-defined architecture in rack-scale clusters and data centers.
\vspace{-2ex}

\end{itemize}

\begin{figure}[t]
        \centering
        %\begin{minipage}[c]{\linewidth}
                \includegraphics[width=0.85\linewidth]{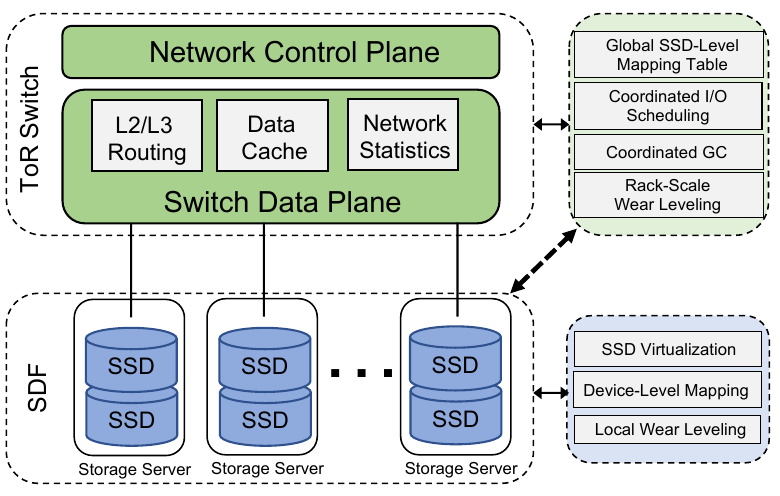}
        %\end{minipage}
        \vspace{-3ex}
        \caption{System overview of \pname{}.}
        \label{fig:overview}
        \vspace{-3ex}
\end{figure}

\subsection{\pname{} Overview}
\label{subsec:overview}
We rethink the software-defined network and storage hierarchy (see Figure~\ref{fig:overview}), and propose a new 
software-defined architecture, \pname{}. 
%Our key insight is that current SDN and SDF stacks lack of
%coordination, which causes suboptimal performance. SDN and SDF share
%similar architectures: 
%the control plane manages programmable hardware devices, 
%and the data plane processes I/O requests. 
%Naturally, to integrate and co-design them, we redefine the functions of their control and data plane to improve the efficiency of the entire rack-scale storage system.
We first decouple the functions of the storage management (i.e., flash translation layer)
of SSDs, and integrate the appropriate functions such as GC and wear-leveling into the SDN ($\S$\ref{subsec:sdf}). 
Such a new architecture enables state sharing between SDN and SDF. It utilizes the capability of SDN to enable 
global storage resource management in a rack.
%After the integration of SDN and SDF, we propose to enable the state sharing of
%I/O events on the data planes of SDN and SDF. Therefore, 
Thus, we can coordinate the efforts of I/O scheduling across the entire rack. 
Henceforth, the SDN and SDF can manage the end-to-end request delay, 
and provide precise feedback to the I/O scheduler on the storage servers ($\S$\ref{subsec:ioschedule}). 
The coordinated I/O scheduling mechanism improves the end-to-end I/O performance and enables intelligent
decision-making in advance. %such as early I/O cancellation or redirection.
To alleviate the performance interference caused by the GC, \pname{} enables coordinated GC  
by exploiting the data replicas in a rack ($\S$\ref{subsec:gc}). As the ToR switch has the global states  
of the SSDs in a rack, it can redirect I/O requests to the corresponding replica upon GC. 
Similarly, \pname{} enables rack-scale wear leveling, as it has the knowledge of the wear of SSDs 
in a rack ($\S$\ref{subsec:wearleveling}). It has a two-level wear leveling mechanism: a local wear balancer 
for ensuring the wear balance in each storage server, and a global wear balancer that reduces 
the wear variance across the entire rack. These wear balancers work at different levels and cooperate 
to ensure rack-scale wear leveling. 
%We now discuss each proposed technique in \pname{} as follows.  

\mbox{\pname{}} manages SSDs at rack-scale for three major reasons. 
First, storage systems are commonly deployed at rack scale, making this a natural granularity for
storage management~\mbox{\cite{www-hadoop-arch, ghemawat:sosp2003}}. Second, rack-scale management is 
facilitated by the programmable ToR-switch with the capability to 
observe the rack's network and I/O traffic. Third, existing rack-aware replica placement schemes make it a natural choice 
for coordinating the GC of SSDs across the rack.
We now discuss each proposed technique in \mbox{\pname{}} as follows. 

\begin{figure}[t]
        \centering
        \includegraphics[width=0.9\linewidth]{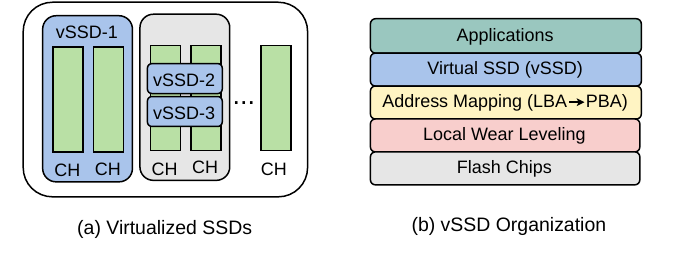}
        \vspace{-3.5ex}
	\caption{The structure of virtualized SSDs. \pname{} supports both 
	hardware-isolated and software-isolated vSSDs.}
        \label{fig:vssd}
        \vspace{-3ex}
\end{figure}

\subsection{Decoupling the Storage Management}
\label{subsec:sdf}

When decoupling storage management, we need to consider two factors: (1) whether integrating an SDF function 
into SDN will benefit from the coordination or not; and (2) if yes, the integration should consume minimum precious 
hardware resources in the switch. We now discuss how
\mbox{\pname{}} decouples storage management between SDF and SDN to maximize the benefits of
co-design while retaining the original flexibility and modularity of SDN and SDF.

\noindent
\textbf{Storage management in SDF.}
%As vSSDs are hosted on each storage server, and 
As the ToR switch has limited hardware resources, 
we keep the essential functions for the vSSD management locally on storage servers (see Figure~\ref{fig:overview}). 
They include SSD virtualization, device-level mapping, and local wear leveling for SSDs in a server. 
%The SSD virtualization is the same as we usually deploy for SDF. 

With SSD virtualization, a programmable SSD can be virtualized into two types of vSSDs: 
hardware-isolated vSSDs, and software-isolated vSSDs. A hardware-isolated vSSD instance is mapped 
to a set of flash channels, as the channel-level parallelism in SSDs provides the strongest 
performance isolation (vSSD-1 in Figure~\ref{fig:vssd}). A software-isolated vSSD is mapped to a set of flash chips, and it 
will share the flash channels with other software-isolated vSSDs, such as vSSD-2 and vSSD-3 shown in Figure~\ref{fig:vssd}. 
It relies on the software-isolation techniques such as token bucket rate limiting to offer relatively weaker 
performance isolation. \pname{} supports both hardware-isolated and software-isolated vSSD instances to 
support different cloud storage services.

%(BEN) want to distinguish difference between wear leveling in Figure 4(b) and wear leveling in 3.6
For each vSSD, it has its own address mapping table (device-level mapping) and local wear leveling 
(i.e., the default wear leveling) for flash block management, as shown in Figure~\ref{fig:vssd}b. 
We keep these functions in the SDF stack, as they are more 
convenient when handled by storage servers. As for other FTL functions, such as bad block management 
and error correction code (ECC) of an SSD, we leave them to the SSD firmware, as the hardware engine 
in SSD controllers is more efficient in managing them.   

%It is worth noting that 
\mbox{\pname{}} enables data replication at vSSD granularity. 
%This is a natural design choice in SDF, as the vSSD abstraction has been shown to simplify the
This is a natural design choice, as the vSSD abstraction has been shown to simplify the
storage management of flash blocks, 
offer flexibility for mapping vSSD instances to underlying flash chips, and incur limited metadata overhead~\mbox{\cite{huang:fast17}}.
%of managing 
%vSSD instances~\mbox{\cite{huang:fast17}}}. 

\begin{figure}[t]
  \centering
  \includegraphics[width=0.6\linewidth]{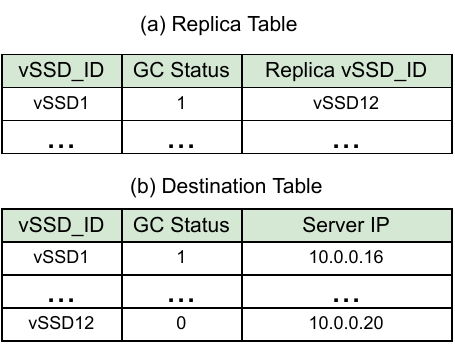}
  \vspace{-1ex}
	\caption{\pname{} tables placed in the ToR switch.}
  \label{fig:switchtable}
  \vspace{-3ex}
\end{figure}

\noindent
\textbf{Storage management in ToR switch.}
To make the ToR switch aware of the states of SSDs in a rack, we maintain a vSSD-level mapping table 
in its data plane, as shown in Figure~\ref{fig:switchtable}. 
%These vSSD states include its GC status and IP address of its replica in the rack. 
%such as the average erase count and the count of write traffic. 
The states tracked in the vSSD-level mapping tables provide the essential knowledge for the 
%coordinated I/O scheduling, coordinated GC, and rack-scale wear leveling. 
coordinated I/O scheduling and coordinated GC. %, and rack-scale wear leveling. 

%To facilitate coordinated I/O scheduling ($\S$\ref{subsec:ioschedule}), 
%\pname{} tracks the network latency for each I/O request to a vSSD, and forwards the 
%measured network latency to the corresponding I/O scheduler in the SDF stack. 
%To enable coordinated GC ($\S$\ref{subsec:gc}), 
%\pname{} uses vSSD-level mapping tables to track the GC status of each vSSD (including 
%its replica vSSD) in a rack. To support rack-scale wear leveling ($\S$\ref{subsec:wearleveling}), 
%\pname{} will track the average wear of each storage server in the data plane of the ToR switch, 
%thus, it can easily obtain the global wear imbalance among storage servers in a rack. 

Specifically, \pname{} maintains two tables in the ToR switch as shown in Figure~\ref{fig:switchtable}: 
\textbf{1)} replica table, which tracks the GC status (1 byte) of each vSSD and its replica vSSD ID (4 bytes); 
\textbf{2)} destination table, which mainly tracks the corresponding server IP (4 bytes) of each vSSD, and the 
GC status (1 byte) of the vSSD. 
%(3) wear table, which tracks the average wear count (4 bytes) of each 
%server in a rack. 
As this mapping table is managed at vSSD granularity,
its storage cost is small, which can be stored in the on-chip memory of the programmable switch.
Given that a rack usually has 64 servers or less, each server has 16 SSDs, and each SSD can be virtualized into 
128 vSSDs, we will have up to 64K vSSDs in a rack\footnote{A typical server in data centers today has 16 PCIe slots, it 
can host 16 SSDs. Assume each SSD has 4TB, the minimum size of a vSSD is 32GB, therefore, each SSD can host up to 128 vSSDs.}. 
The maximum size of each table is 1.3MB. The total size of these tables for \pname{} is much less than the available SRAM capacity 
%(at least 4MB) in modern programmable switches. 
(tens of MBs) in modern programmable switches.

\begin{figure}[t]
  \centering
  \includegraphics[width=0.95\linewidth]{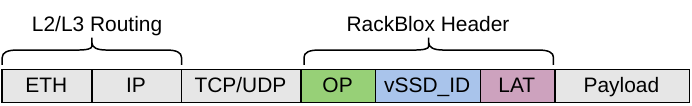}
  \vspace{-1ex}
  \caption{The network packet format in \pname{}.}
  \label{fig:packetformat}
  \vspace{-2ex}
\end{figure}

\noindent
%\textbf{State communication between ToR switch and storage servers.}
\textbf{State communication between SDN and SDF.}
To facilitate the state communication between the ToR switch and storage servers in the same rack, 
\pname{} leverages the programmability of SDN, and has its own network packet format based on regular network protocols, as shown 
in Figure~\ref{fig:packetformat}. The packet has one 1-byte \textit{OP} field to indicate 
different operations as shown in Table~\ref{tab:packetop}, one 4-byte field to indicate the 
target vSSD ID, and one 4-byte field (\textit{LAT}) for storing the measured network latency for the  
packets transferred through the data center network. The payload will be filled with different values, according to 
the operation specified in \pname{} header. 
%TODO CUT Candidate
We will discuss the purpose of each operation throughout the paper. 

The \mbox{\pname{}} header is part of the L4 payload. \mbox{\pname{}} uses existing L2/L3 routing protocols to
route packets. As such, switches can forward \mbox{\pname{}} packets normally, and 
\mbox{\pname{}} is compatible with flow/congestion control and other network functionalities in the
transport layer. 
We differentiate \mbox{\pname{}} packets in the ToR switch through a reserved TCP/UDP port. 
%while other packets are forwarded normally.

To initialize \pname{} tables in the ToR switch, the storage servers send a packet  
that contains the \textit{create\_vssd} operation to the switch upon creating a new vSSD. 
The \textit{vSSD\_ID} field will store the ID of the newly created vSSD, the payload will include 
the 4-byte server IP, its replica vSSD ID, and the server IP of the replica vSSD.  
%(BEN) Feels a bit redundant, but it confused one reviewer so this is more direct
The replica vSSD ID and IP are allocated with the vSSD--following the rack-aware replica
placement scheme in rack-scale storage systems. The ToR 
%switch will insert a new entry in the replica table and destination table, respectively with 
switch will insert a new entry in the replica table and destination table, with 
GC states initialized as 0 (idle). Upon vSSD deletion, the storage server 
sends the \textit{del\_vssd} packet to the ToR switch to remove the corresponding entries in the \pname{} tables. 
As we serve I/O requests at runtime, \pname{} tables will be updated depending on 
the events. 
%in the rack (see the remaining sections). %We will discuss how they are updated in the remaining sections.  

\begin{table}[t]
    \centering
    \scriptsize
	\caption{Network protocols used in \pname{}.}
    \vspace{-3.5ex}
    \label{tab:packetop}
    %\begin{tabular}{ |c|c|c|c|c|} 
        \begin{tabular}{|p{70pt}<{\raggedleft}|p{140pt}<{\raggedright}|}
            \hline
            \textbf{Operation Name} & \textbf{Description} \\
            \hline
            \textbf{\texttt{create\_vssd}} & Register a newly created vSSD in the ToR switch. \\
            \hline
            \textbf{\texttt{del\_vssd}} & Remove a registered vSSD from the tables.\\
            \hline
            \textbf{\texttt{write}} & Write issued by client. \\
            \hline
            %\textbf{\texttt{Write-Reply}} & Server reply packet for write\\
            %\hline
            \textbf{\texttt{read}} & Read issued by client. \\
            \hline
            %\textbf{\texttt{Read-Reply}} & Server reply packet for read \\
            %\hline
            \textbf{\texttt{gc\_op}} & Packet to update GC for vSSD. \\
            \hline
        \end{tabular}
	    \vspace{-3.5ex}
\end{table}

    \vspace{-1ex}
\subsection{Coordinated I/O Scheduling}
\label{subsec:ioschedule}
%\paragraph{Research Task 2: Coordinated I/O scheduling for predictable end-to-end performance.}
%As discussed, the incoordination between SDN and SDF 
%makes it hard to guarantee constant low end-to-end latency for networked I/O requests. 
Although both SDN and SDF can make the best effort to achieve their quality of service (QoS), 
the lack of state sharing and coordination will cause suboptimal end-to-end performance and 
wasted effort on I/O scheduling. %(which further causes resource inefficiency). 
%%their scheduling policies may conflict with each other due to the incoordination.
%For instance, a network operator could use the Least Slack Time First policy to minimize the tail packet
%delays~\cite{slack:algorithm, pifo:sigcomm2016}. However, the storage stack widely uses
%the weighted fair-sharing mechanism~\cite{ahn:hotstorage2016} for fair sharing, without considering
%the priority of requests having the least time slack.
%This  
%could still cause long tail latency due to the interference between multi-tenant
%applications~\cite{huang:fast17}.
%Second, it can cause wasted efforts on I/O scheduling, which further
%results in resource wastage and traffic congestion.
For instance, as SDN and SDF are independent, they do not share states of I/O requests, therefore, 
%SDN may forward I/O requests to busy storage servers anyway, although it is apparent
SDN may forward I/O requests to busy storage servers, although it is apparent
that their end-to-end service-level objective is likely to be violated. This will exacerbate 
network congestion and increase the pressure of processing I/O requests on storage servers. 

\pname{} enables the state sharing of I/O requests across SDN and SDF, and 
%develop a coordinated mechanism that SDN and SDF can precisely predicate the
%end-to-end delay of I/O requests, and 
develops a coordinated I/O scheduling mechanism to improve end-to-end performance. 
It tracks the elapsed time in the programmable switches, adapts the I/O scheduling in the data
plane of SDF to control the end-to-end delay for the request, and predicts the time it would take
to transmit the response from the storage server to the client.

%Given a desired end-to-end SLO, 
\pname{} tracks I/O requests %to obtain this knowledge 
across the entire stack:
(1) $Net_{time}$: the elapsed time in the network stack since the I/O request is issued until it reaches the storage server; 
(2) $Storage_{time}$: the delayed time in the I/O queue of the storage stack; 
(3) $Predict_{time}$: the time it takes to transfer the response back to the client 
over the network. 
To manage I/O scheduling in SDF, \pname{} uses 
%$Prio_{sched}$ = ($Net_{time}$ + $Storage_{time} + $ $Predict_{time}$) as an indicator of the scheduling priority. 
$Prio_{sched}$ = ($Net_{time}$ + $Storage_{time} + $ $Predict_{time}$) as the scheduling priority. 
As \pname{} issues I/O requests from the queue 
in the storage stack, it selects the request with the maximum $Prio_{sched}$ value. 
%Different from state-of-the-art storage I/O scheduling schemes (e.g., priority-based~\cite{cfq} 
%and deadline-based~\cite{deadline}), \pname{} considers the network 
%latency and makes the best effort to reduce the end-to-end latency. 
\pname{} differs from state-of-the-art storage I/O scheduling schemes by considering the network latency to make the best effort to reduce
the end-to-end latency~\cite{deadline, cfq}.

%\begin{table}[t]
%    \centering
%    \scriptsize
%	\caption{Network protocols used in \pname{}.}
%    \vspace{-3.0ex}
%    \label{tab:predacc}
%    %\begin{tabular}{ |c|c|c|c|c|} 
%        \begin{tabular}{|p{70pt}<{\centering}|p{35pt}<{\centering}|p{35pt}<{\centering}|}
%            \hline
%            \textbf{Network Distribution} & \textbf{Accuracy} & \textbf{Error} \\
%            \hline
%            \textbf{Fast}~\cite{iqbal:ton2022} & 99.9\% & 1.09\% \\
%            \hline
%            \textbf{Medium}~\cite{mogul:hotos2015} & 99.8\% & 0.58\% \\
%            \hline
%            \textbf{Slow}~\cite{popescu:tma2018} & 86.6\% & 9.82\% \\
%            \hline
%        \end{tabular}
%	    \vspace{-4.0ex}
%\end{table}

In order to track $Net_{time}$ with low overhead, we use 
the In-band Network Telemetry (INT) available in programmable switches~\cite{int}. 
It enables the network state collection in the data plane without intervention from the 
%control plane. \pname{} uses INT to compute the sum of per-hop latency, since the 
%(BEN) clarify a bit
    control plane. \pname{} uses INT to compute the sum of per-hop latency in the switches, since the 
routing and queuing latencies dominate the network latency~\cite{telemetry,pingmesh:sigcomm2015}. 
It embeds the measured network latency in the network packet being transferred to 
the storage server, following the network format in Figure~\ref{fig:packetformat}.
As for $Storage_{time}$, \pname{} tracks the queuing delay for each I/O request 
in the queue of the storage stack. 
%and predict the latency of storage I/O request with SDF.

To predict the time it will take to return the response to the 
client ($Predict_{time}$), we develop a predictor using a simple yet effective sliding window algorithm. 
%as they are efficient in time-series predictions. 
We track one sliding window for each vSSD with the average network latency of the $100$ most
%(BEN) clarify that it is *incoming* packets
    recent incoming packets. We choose 100 packets because it is small enough to quickly detect changes in the
network (e.g., network congestion), but large enough to smoothen outlier requests. We use 
    incoming packets because they can better capture the factors causing network delays. 
    %than the more stale data returned from outgoing packet latencies.
We maintain separate
windows for reads and writes as their outgoing packet sizes are different%, a factor in serialization delay
~\mbox{\cite{timecard:sosp2013,timely:sigcomm2015,prioritymeister:socc2014}}. 

Our experiments with a variety
    of network traces in data centers~\mbox{\cite{iqbal:ton2022,mogul:hotos2015,popescu:tma2018}}
    (see $\S$\mbox{\ref{subsec:implt}} for details) show that this
approach effectively predicts the return latency. The predicted latency is
within 25$\mu$s of the correct value 95\% of the time (across all distributions) and 
86.6\% of the time in the worst case. %The magnitude of mispredictions is critical, 
The predictions are within 10\% of the true latency in the
worst case. 
    Mispredictions primarily occur at the begin/end of
    congestion or with highly variables network patterns. 
We show the benefit of the coordinated I/O scheduling in $\S$\mbox{\ref{sec:eval}}.

%% GC Section is in separate file
\vspace{-2.5ex}
\subsection{Coordinated Garbage Collection}
\label{subsec:gc}

The GC overhead of SSDs is significant, as it blocks incoming I/O requests and seriously harms end-to-end latency~\cite{Maas-asplos16, Ousterhout-nsdi15}. 
For instance, a 4KB read request in SSDs can be completed in under 100 $\mu$s, but it may wait for a few milliseconds 
due to the GC. This is critical in data centers, where many applications have strict performance requirements~\cite{microsecond:barroso}.

The fundamental issue is that the upper-level system software fails to consider the underlying SSD behavior. 
Without coordination at rack scale, it is hard to optimize GC across replicas or redirect requests away
from replicas executing GC, even though SDF exposes the underlying storage behavior.
Since the ToR switch will forward each storage request entering the rack, it has the states of the SSDs in the rack, 
%it is natural to utilize the switch to coordinate the GC events across these SSDs. 
it is natural to coordinate GC across the SSDs with the switch. 

Prior work has explored various techniques for coordinating GC between SSDs within
servers~\mbox{\cite{li:sosp2021,yan:fast2017,kim:msst2011,kim:atc2019,skourtis:atc2014,shin:fast2013,kim:toc2014}}.
These studies managed SSDs either spatially by reserving spare SSDs to serve requests or temporally by
scheduling GC to ensure predictable latency for read requests. 
However, they did not enable GC coordination across servers at rack scale. 
Industry has been developing rack-scale storage solutions~\mbox{\cite{hitachivantara,purestorage,silk,purity:sigmod2015}}, however, 
to the best of our knowledge, they also lack GC coordination across data replicas.

As different levels of vSSD isolation (software vs. hardware) have different challenges, we will begin with coordinating GC 
for hardware-isolated vSSD instances and then extend it to software-isolated vSSDs.

\begin{figure}[t]
    \centering
    \includegraphics[width=0.93\linewidth]{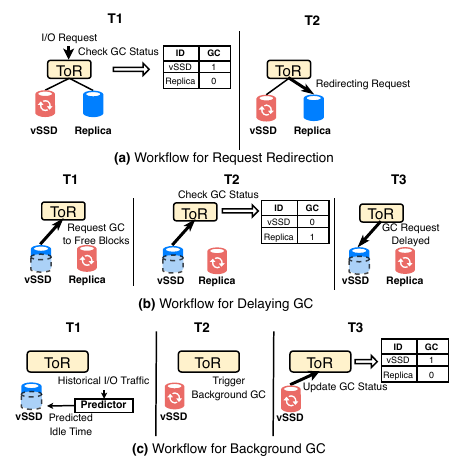}
    \vspace{-4ex}
    \caption{Coordinated GC optimizations in \pname{}.}
    \label{fig:gc-overview}
    \vspace{-3ex}
  \end{figure}
\subsubsection{Coordinated GC for Hardware-Isolated vSSDs}
\label{subsubsec:hw_gc}
\mbox{\pname{}} coordinates GC between the replicas of each vSSD, as shown in Figure~\ref{fig:gc-overview}.
Since each hardware-isolated vSSD is mapped to one or more flash channels that run GC independently, 
GC in other vSSDs does not affect its performance. 
%Thus, by coordinating GC and tracking shared states per vSSD, we only need to coordinate GC events  
%between its replicas. 
%Thus, by coordinating GC and tracking shared states per vSSD, we can achieve predictable performance.
Thus, by coordinating GC per vSSD, we can achieve predictable performance.
Coordinating GC at vSSD granularity also aligns with the granularity of data replication. If we
keep at least one vSSD replica idle, we can ensure that one data replica can be accessed with predictable
performance. 
%not between physical SSDs or replicas of other vSSDs. 

%We show the coordinated GC workflows in . %and explain them below.
%For each vSSD, we achieve predictable performance for storage requests by enabling request redirection. 
For each vSSD, we achieve predictable performance for storage requests by enabling request redirection (Figure~\ref{fig:gc-overview}a). 
While requests are directed to idle replicas, 
other replicas of the vSSD may still run GC.  
Thus, the switch delays GC for one replica until another is ready to serve requests (Figure~\ref{fig:gc-overview}b). 
Also, \pname{} enables background GC to utilize the idle cycles of SSDs (Figure~\ref{fig:gc-overview}c).

We outline the packet processing workflow of SDN and SDF in Algorithm~\ref{algo:sdn} and Algorithm~\ref{algo:sdf}. 
%We examine \pname{} with two replicas per vSSD in a rack, following  
%the common settings of rack-scale storage systems. 
We examine \pname{} with two replicas in a single rack and 
one replica in another following the common rack-scale storage systems~\cite{ghemawat:sosp2003,www-hadoop-arch,www-kafka,www-hadoop-block-placement,miao:sigcomm2022}.
\pname{} can be extended to any number of replicas.

\noindent
\textbf{Request Redirection.}
Upon receiving a packet, \pname{} queries the Replica Table (see $\S$\ref{subsec:sdf}) to get the
\textit{gc\_status} and \textit{replica} for \textit{vssd\_id}. 
Writes are not redirected (Line 2-3 in Algorithm~\ref{algo:sdn}), but issued to all replicas for reliability and
consistency~\cite{www-hadoop-arch,ghemawat:sosp2003,miao:sigcomm2022,gao:nsdi2021}.
\mbox{\pname{}} supports different consistency models, and our implementation uses Hermes~\mbox{\cite{hermes:asplos2020}} to
ensure strong consistency between replicas and correctness
when redirecting requests.
We avoid long tail latencies for writes by utilizing existing DRAM caches in data center servers to absorb
writes during GC~\cite{gao:nsdi2021, li:sosp2021}. This follows the durability semantics
of existing systems that primarily rely on replicas to ensure data
durability~\mbox{\cite{ramcloud1, pangu:fast2023}}. Writes are considered complete when all
replicas have a DRAM copy and are flushed in the background (Line
2-4 in Algorithm~\ref{algo:sdf}).

For read requests, if the \textit{gc\_status} is set for the \textit{vssd\_id}, we query the
Destination Table to get the
\textit{gc\_status} for the replica vSSD (T1 in Figure~\ref{fig:gc-overview}a). If the vSSD is not executing GC or if both
the vSSD and its replica are executing GC, we forward the packet as is. Otherwise, 
%we redirect the request to the replica vSSD using its destination IP obtained from the Destination Table 
we redirect the request to the replica using its destination IP in the Destination Table 
(T2 in Figure~\ref{fig:gc-overview}a, Line 4-9 in Algorithm~\ref{algo:sdn}). 
%The received I/O requests are scheduled locally with coordinated I/O scheduling (Line 5-6 in
I/O requests are scheduled locally with coordinated I/O scheduling (Line 5-6 in
%Algorithm~\ref{algo:sdf}). By redirecting requests, \pname{} ensures that requests are served by 
Algorithm~\ref{algo:sdf}). By redirecting requests, requests are served by 
replicas without suffering from the GC overhead.

While \pname{} maximizes the chance that at least one replica is available, it is
possible that both replicas are executing GC. The techniques that submit requests to another 
rack in parallel can be applied to ensure high performance~\cite{tailatscale}. In this paper, 
we focus on the intra-rack I/O scheduling. 
%\hl{but \mbox{\pname{}} can be extended to multiple switches
%by modifying the the GC logic in Algorithm~\mbox{\ref{algo:sdn}} to forward GC update packets to
%other switches and keep the GC states consistent.}
%and we will explore the inter-rack scheduling as future work. 

\newcommand\mycommfont[1]{\footnotesize\ttfamily\textcolor{purple}{#1}}
\SetCommentSty{mycommfont}
% \setlength{\textfloatsep}{7pt}
% \RestyleAlgo{boxruled}
\SetAlFnt{\small}
\begin{algorithm}[t]
    \DontPrintSemicolon % Some LaTeX compilers require you to use \dontprintsemicolon instead
    \SetAlgoNoEnd
    \SetAlgoLined
    \KwIn{$pkt \leftarrow$ \pname{} packet \\ 
    \ \ \ \ \ \ \ \ \ \ \ \ $gc\_status\leftarrow$ GC status\\ 
    \ \ \ \ \ \ \ \ \ \ \ \ $dst \leftarrow$ table mapping vssd\_id to its destination IP\\
    \ \ \ \ \ \ \ \ \ \ \ \ $replica \leftarrow$ replica for this vssd\_id}
    %\tcp{\color{purple}\textbf{Packet Processing for Coordinated GC}}
    \SetKwFunction{FMain}{$process\_packet$}
    \SetKwProg{Fn}{Function}{:}{}
    \Fn{\FMain{$pkt,\ gc\_status,\ replica$}}{
        {\If{$pkt.op$ = $write$}{
            $forward$($pkt$) \; 
        }}
        {\If{$pkt.op$ = $read$}{
            {\If{$gc\_status[pkt.vssd\_id]$ = 1}{
                {\If{$gc\_status[replica]$ = 0}{
                    $pkt.dst \leftarrow dst[replica]$\;
                    $pkt.vssd\_id \leftarrow$ $replica$\;
                }}
            }}
            $forward$($pkt$) \; 
        }}
        {\If{$pkt.op$ = $gc\_op$}{
            $gc\_status[pkt.vssd\_id] \leftarrow 1$\;
            {\If{$pkt.gc$ = $soft$}{
            % {\If{$pkt.gc$ = $soft\ \textbf{or}\ pkt.gc$ = $bg$}{
                \tcp{\color{purple}\textbf{requires recirculation}}
                {\If{$gc\_status[replica]$ = $1$}{
                    $pkt.gc \leftarrow\ delay$\;
                    $gc\_status[pkt.vssd\_id] \leftarrow 0$\;
                }}
                {\Else{
                    $pkt.gc \leftarrow\ accept$\;
                    $dst\_gc\_status[pkt.vssd\_id] \leftarrow 1$\;
                }}
            }}
            {\ElseIf{$pkt.gc$ = $finish$}{
                $gc\_status[pkt.vssd\_id] \leftarrow 0$\;
            }}
            {\Else{
                $dst\_gc\_status[pkt.vssd\_id] \leftarrow 1$\;
                $pkt.gc \leftarrow\ accept$\;
            }}
            $pkt.dst \leftarrow\ pkt.src$\;
            $forward$($pkt$) \; 
        }}
    }
    \caption{{\sc \pname{} workflow in SDN}}
    \label{algo:sdn}
\end{algorithm}                                       

\begin{algorithm}[t]
    \DontPrintSemicolon % Some LaTeX compilers require you to use \dontprintsemicolon instead
    \SetAlgoNoEnd
    \SetAlgoLined
    %\tcp{\color{purple}\textbf{Packet Processing}}
    \KwIn{$pkt \leftarrow\ $\pname{}\ packet}
    \SetKwFunction{FMain}{$process\_packet$}
    \SetKwProg{Fn}{Function}{:}{}
    \Fn{\FMain{$pkt$}}{
        {\If{$pkt.op$ = $write$}{
            {\If{$cache$ is full}{
                flush DRAM cache with write data
            }}
        }}
        {\If{$pkt.op$ = $read$}{
            schedule local read with coordinated I/O\;
        }}
        {\If{$pkt.op$ = $gc\_op\ \textbf{and}\ pkt.gc$ = $accept$}{
            begin GC\;
        }}
    }
    \tcp{\color{purple}\textbf{Periodic GC Monitoring}}
    \KwIn{$vssd \leftarrow\ $vSSD being checked for GC
    \ \ \ \ \ \ \ \ \ \ \ \ $soft\_threshold \leftarrow$ soft GC threshold\\ 
    \ \ \ \ \ \ \ \ \ \ \ \ $gc\_threshold \leftarrow$ regular GC threshold\\
    \ \ \ \ \ \ \ \ \ \ \ \ $bg\_pred \leftarrow$ idle prediction for background GC}
    \SetKwFunction{FMain}{$trigger\_gc$}
    \SetKwProg{Fn}{Function}{:}{}
    \Fn{\FMain{$vssd,bg\_pred$}}{
        $gc\_type \leftarrow\ none$\;
        {\If{$vssd.free\_blocks$ < $gc\_threshold$}{
            $gc\_type \leftarrow\ regular$\;
        }}
        {\ElseIf{$vssd.free\_blocks$ < $soft\_threshold$}{
            $gc\_type \leftarrow\ soft$\;
        }}
        {\ElseIf{$bg\_pred$ = $True$}{
            $gc\_type \leftarrow\ bg$\;
        }}
        {\If{$gc\_type \neq none$}{
            $pkt \leftarrow\ \textbf{new}$ $pkt$\;
            $pkt.dst,\ pkt.op,\ pkt.gc \leftarrow\ switch,\ gc\_op,\ gc\_type$\;
        }}
    }
    \caption{{\sc \pname{} workflow in SDF}}
    \label{algo:sdf}
\end{algorithm}

\noindent
\textbf{Delaying GC.}
Since all replicas receive the same writes, replicas may  
execute GC at the same time~\cite{li:sosp2021}. Therefore, naive request
redirection cannot alleviate the GC overhead. 
%that has storage servers notify the switch of their GC status without coordination of replicas 
%cannot easily decouple their GC. 
To overcome this issue, we leverage the shared states in the
switch and empower the switch to delay the GC of a replica. %if GC windows overlap. 

While delaying GC can ensure that two replicas do not execute their GC simultaneously, we cannot
delay indefinitely. GC is typically executed when the available free blocks fall below a fixed
\textit{gc\_threshold} (e.g., 25\%). 
%and terminates when a sufficient number of blocks have been erased. 
This is a hard threshold to free flash blocks for future use. 
%enable internal storage events such as bad block management. 
To make room for delaying GC, we configure a relaxed
\textit{soft\_threshold} (35\% by default). Instead of
having the SDF notify the switch when it must do GC, it requests GC once its free block
ratio falls below the \textit{soft\_threshold}. The switch can use its
shared GC state to delay GC until the replica finishes GC.

%To trigger GC in \pname, 
Storage servers will periodically (every $30$ seconds by default) check the free block ratio
for each vSSD {(Line 9-19 in Algorithm~\mbox{\ref{algo:sdf})}}. If any GC condition triggers, the SDF
will send a \textit{gc\_op} packet (T1 in Figure~\mbox{\ref{fig:gc-overview}}b).
If the free block ratio falls below the \textit{gc\_threshold}, the 
\textit{gc} field in the payload is set to \textit{regular} (value of 1) to indicate that the replica must
execute GC. GC requests
with \textit{regular} will not be denied as the GC has been delayed as much as possible. 
If \textit{regular} GC requests are not acknowledged due to link or
switch failure, the vSSD will execute GC after retrying (3 retries by default).
If the free block ratio only falls below the \textit{soft\_threshold}, the \textit{gc} field in the payload is set to \textit{soft} (value of 0).

The logic for \textit{accepting} or \textit{delaying} GC requests in the switch is shown in
Line 10-25 of Algorithm~\ref{algo:sdn}. The switch begins with updating the GC status in the Replica Table to $1$. 
If the request is \textit{regular}, the switch also updates the GC status in the Destination Table to $1$,
sets the \textit{gc} field in the payload to \textit{accept} (value of 3), and sends
the reply back to the server. 
For \textit{soft} requests, the switch will check the GC status of the
replica {(T2 in Figure~\mbox{\ref{fig:gc-overview}}b)}. If the replica is executing GC, 
the switch will \textit{delay} (value of 4) the request {(T3 in Figure~\mbox{\ref{fig:gc-overview}}b)}.
Otherwise, the switch will \textit{accept} it. Both the Replica and Destination Tables have a
GC status that must be
consistent. The \textit{soft} requests that must check the replica's GC status in the Destination Table 
cannot also update the GC status of the vSSD due to the memory limitations of programmable switches. 
Therefore, we recirculate the packet once to ensure consistency. %In practice, GC updates are very infrequent, making recirculation a minor overhead. 
The SDF sends a final \textit{gc\_op} packet when the vSSD has finished GC with the \textit{gc} field
set to \textit{finish} (value of 5) in the payload. The switch uses this to clear the GC status in 
both tables (Line 19-20 in Algorithm~\ref{algo:sdn}).

\noindent
\textbf{Background GC.}
Delaying GC enables the switch to reduce overlapping GC.
%However, worst case workloads may trigger GC intensively,
%leading to overlapping GC anyways. The best option is to 
\pname{} also opportunistically utilizes idle cycles to free blocks. 
%To that end, we integrate background GC into
%\pname{}.
Background GC requests are labeled as \textit{bg} (value of 2) in the \textit{gc} field of the payload. 
Since background GC is performed during idle cycles, the SDF executes it without approval from the
ToR switch. 
 %does not require approval from the ToR switch to run it. 
%The switch treats the background GC requests as \textit{regular} requests as they both cannot be delayed (Line 21 of Algorithm~\mbox{\ref{algo:sdn}}).
To facilitate background GC, % in the storage server, 
\pname{} predicts the next idle time for a given vSSD based on the last interval between I/O requests~\cite{eitan:sigmetrics2002, mi:storage, jian:eurosys2019, rssd:asplos2022}, 
as shown in T1 in Figure~\mbox{\ref{fig:gc-overview}}c:  
$\label{equ:idle}
T_{i}^{predict} = \alpha * T_{i-1}^{real} + (1 - \alpha) * T_{i-1}^{predict}$, where $T_{i-1}^{predict}$ is the idle time of the last prediction, 
and $\alpha$ is the exponential smoothing parameter ($\alpha=0.5$ by default).
Once $T_{i}^{predict}$ is larger than a defined threshold (30 milliseconds by default), the storage server will execute GC and update the GC status 
in the switch (T2 and T3 in Figure~\mbox{\ref{fig:gc-overview}c}).

\subsubsection{Coordinated GC for Software-Isolated vSSDs}
\label{subsubsec:sw_gc}

Unlike hardware-isolated vSSDs, 
software-isolated vSSDs can share channels with other software-isolated vSSDs. 
As they rely on software techniques to offer performance isolation, 
software-isolated vSSDs provide relaxed isolation guarantees.  
Thus, request redirection may not guarantee
predictable storage performance for those vSSDs.
Even if one replica is not executing GC, a collocated vSSD may
execute GC, resulting in significant interference. 

%%One paragraph to explain the background harvesting how this interacts with the switch.
\pname{} enables simple management of software-isolated vSSDs by grouping them into channel groups
in the SDF. Each channel group is a set of software-isolated vSSDs that span the same set of
channels and all vSSDs of the channel group will perform GC simultaneously. Intuitively, if one
vSSD must perform GC and each vSSD will be affected anyway, then all vSSDs should
perform GC to reduce GC frequency. This simplifies coordination and reduces overhead. 

%%One paragraph to describe how we make this make sense.
The channel group is managed exclusively by the SDF and is not exposed to the switch. 
%However, an individual vSSD may consume all of its free blocks before the channel reaches its
%\textit{soft\_threshold} and may need to free blocks. To avoid such conflicts, 
Multiple software-isolated vSSDs sharing the same channels may have diverse GC behavior. 
%To minimize their impact on each other, we allow one vSSD that requires GC to 
To ensure all vSSDs of the channel group can execute GC together, we allow a vSSD that has
exhausted its free blocks to transparently borrow free blocks from collocated vSSDs.
Blocks are borrowed in groups (1GB by default) and transferred between the free block lists of
vSSDs.
Thus, we can delay GC until the channel group's free block ratio falls below the
\textit{gc\_threshold}. The borrowed blocks will be erased (for security) and returned to the original vSSD after the GC. 
The coordinated GC will not worsen the write amplification, as it makes the best effort to avoid unnecessary GC operations. 
%(BEN) clarify concerns about what thresholds are used and if this has write amplification issues
%\hl{GC is initiated for channel groups using the same thresholds used in
%$\S$\mbox{\ref{subsubsec:hw_gc}}. By using the same thresholds as hardware-isolated vSSDs,
%channel groups can help alleviate potential write-amplification concerns for GC in software-isolated vSSDs.}
When sending \textit{gc\_op} packets to the
switch, the storage server generates a separate packet for each vSSD in the channel group. Note that a
\textit{delay} response (i.e., the corresponding replica vSSD is executing GC) from any vSSD 
will delay the GC of the channel group.

\subsection{Rack-Scale Wear Leveling}
\label{subsec:wearleveling}

The limitation
of the SSD lifetime has created complexity for their use and management in practice~\cite{maneas:fast2022, maneas:fast2020}.
This is especially true in large-scale data centers. 
First, as different applications have different workload patterns,
the write traffic to each SSD can be different, causing wear imbalance between SSDs in a rack. 
%For example, different application has different workload patterns.
%Our study with a variety of data center workloads demonstrates how various they erase blocks
%at different rates, as shown in Figure~\ref{fig:wear}.
%(2) not all SSDs are produced equal, they may have variations in their endurance and reliability;
Second, platform operators have to replace unhealthy or failed SSDs with new SSDs frequently,
making the wear management of SSDs across the entire rack even harder. 
Third, modern cloud infrastructures mostly consider the load balance 
rather than the wear balance across SSDs. 
%were designed without considering the low-level SSD endurance. 
Therefore, the wear-leveling
management of SSDs has become a fundamental challenge in data centers today. 
%due to the tussle between these three issues discussed above. 
%This not only increases the operation cost, but also affects
%the storage performance.

\begin{figure}[t]
  \centering
  \includegraphics[width=0.9\linewidth]{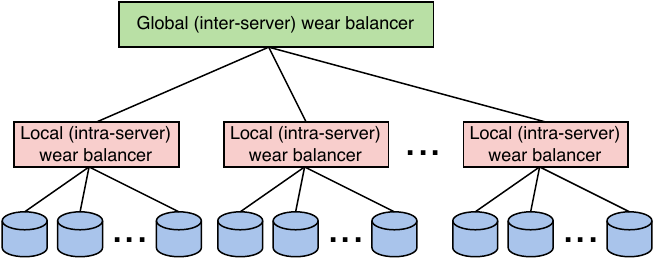}
  \vspace{-2ex}
  \caption{Two-level rack-scale wear leveling in \pname{}.}
  \label{fig:rackscale-wear}
  \vspace{-2.5ex}
\end{figure}

%    Furthermore,} 
Premature death or removal of an SSD not only increases the operation cost, but causes an opportunity loss of other hardware components,
%Such an imbalance in capability of storage servers represents lost opportunity costs,
given that others like CPU, network, and memory do not prematurely lose capability.
%Moreover, unpredictable changes in
%capabilities also complicate the jobs of load balancers that typically assume uniform or predictably
%non-uniform (by design) capabilities.
Thus, it is desirable to ensure SSDs in a rack are aging at the same rate.

To extend the lifetime of a rack of SSDs, we propose a two-level wear leveling mechanism (see Figure~\ref{fig:rackscale-wear}). 
%which balances wear within each individual SSD in a storage server as well as across SSDs in the rack. 
It consists of two parts: a local (intra-server) wear balancer processes the local wear balance between
SSDs in a storage server, and a global (inter-server) wear balancer reduces the wear variance for SSDs in a rack.
The wear balancers work at different level and cooperate to ensure rack-scale wear leveling.

As for the local wear balancer, to obtain the uniform lifetime among SSDs in a storage server, we
track the average erase count for an SSD\footnote{For the programmable SSDs, we can track the erase
count of each flash block. Therefore, we can obtain the average erase count of a flash channel as well as an SSD.},
and ensure the wear balance for all the SSDs in a storage server.
Let $\varphi_{i}$ denote the wear (average erase count of all the blocks to date) of the
$i^{th}$ SSD. $\lambda$ = $\varphi_{max} / \varphi_{avg}$ denotes the
wear imbalance across SSDs,  which must
not exceed 1 + $\gamma$, where $\varphi_{max} = Max(\varphi_{1}, ...,
\varphi_{N})$, $\varphi_{avg} = Avg(\varphi_{1}, ..., \varphi_{N})$, $N$
is the total number of SSDs, and $\gamma$ represents the maximum permitted imbalance.
%We use the ratio of maximum to
%average to quantify the wear imbalance, as the lifetime of ideal wear-leveling
%is determined by the average wear of all the SSDs.
%If $\lambda$ exceeds 1 + $\gamma$, we need to swap SSDs to balance the wear of all the
%SSDs in a server. Instead of swapping SSDs frequently, \pname{} periodically
%(BEN) clarify why SSD swapping helps alleviate wear
%If $\lambda$ exceeds 1 + $\gamma$, \hl{we need to swap which vSSDs write to which SSDs to balance the wear of all the SSDs in a server.} 
Instead of swapping SSDs frequently, \pname{} periodically
swaps the SSD that has incurred the maximum wear with the SSD that has the minimum rate of wear,
following the relaxed wear leveling approach developed in~\cite{huang:fast17}. Given 
$\gamma$ = 0.1, each server can have 16 SSDs, and each SSD can last five years, \pname{} 
can achieve uniform lifetime for SSDs in a storage server by swapping once per 12 days 
for the worst case~\cite{huang:fast17}. 
Assume each flash block can endure 30K writes, this swapping consumes only 0.5\% of its lifetime. 

SSDs are swapped atomically by pausing operations for the chosen blocks, reading them into memory, writing them
%to their new locations, updating the mapping tables accordingly, and serving the paused requests. 
to their new locations, updating the mapping tables, and serving the paused requests. 
As the swapping occurs infrequently,
%and has lower priority than regular I/O requests, 
it does not affect the tail latency. 
%As swapping an SSD will take about 14 minutes, the overall performance 
%%(BEN) add maximum perf degradation
%drops by at most 35\% for only for 0.08\% of all the time and will not affect the tail latency of
%applications~\mbox{\cite{huang:fast17}}.} 
To further minimize its impact on application performance, \pname{} 
assigns higher priority to regular I/O requests during swapping.  
%But how frequently we should swap SSDs in a storage server is bounded by the worse case (a write-only
%workload targets a single SSD).

%Previous study~\cite{huang:fast17} on channel-level swapping in an SSD
%found that we can reach ideal SSD lifetime by swapping channels once per 19 days for datacenter workloads.
%This implies that a relaxed wear leveling scheme will be sufficient to achieve almost
%ideal wear balance between SSDs in a storage server. To minimize the negative impact of the
%SSD swapping, we can redirect the I/O requests to their data replicas during the swapping.

%Similarly, the global wear balancer monitors the average erase count of SSDs in a storage server. 
Similarly, we can quantify the wear imbalance between storage servers
in a rack by using the wear (average erase count of all the SSDs to date) of a server. 
However, different from the swapping of SSDs in a single server, the swapping cost between  
%(BEN) add that in-flight requests may need to be redirected
storage servers is more expensive, due to the networking overhead. %\hl{and potential to need to redirect requests}. 
Therefore, we relax the swapping 
frequency (8 weeks by default). This is less of a concern, as modern storage infrastructures have employed  
the load balance (e.g., round-robin vSSD allocation) across servers. 
%(BEN) resolve confusion about whether wear leveling is done in the switch
Since \mbox{\pname{}} does not swap SSDs across servers frequently, we do not implement the rack-scale wear 
leveling in the switch to keep our design simple. 
Our experiments (see $\S$\ref{subsec:eval_wear}) 
show that the relaxed wear leveling will ensure near-ideal wear balance for datacenter workloads.

\subsection{Implementation Details}
\label{subsec:implt}

\noindent
\textbf{Testbed.}
Our experiments are conducted on a testbed of five servers connected to a 6.5Tbps Tofino
switch~\cite{edgecore_switch}. Each server is equipped with a 24-core Intel Xeon E5-2687W processor 
running at 3.00GHz, 108GB DRAM, and 1TB programmable SSD. %Each server runs Ubuntu Linux 18.04. To connect the switch, 
Each server has a Mellanox ConnectX-4 50Gb NIC connected to the programmable switch. 

\noindent
\textbf{Network implementation.}
To implement the custom packets described in Figure~\ref{fig:packetformat}, we use DPDK
(v22.11.1)~\cite{dpdk}.  
%The \emph{OP}, \emph{vSSD\_ID}, and \emph{LAT} header fields add 1, 4, and
%4 bytes respectively to every packet header. 
If the packet type is \textit{gc\_op}, the
payload contains a \emph{gc} field (1 byte) storing the necessary type of GC request. When the
packet is a \textit{create\_vssd} packet, we include the server IP, vSSD\_ID,
and server IP of the replica in the payload. %This adds three headers that are $4$ bytes each. 

We develop the switch data plane in P4~\cite{pat:sigcomm2014} and run it on an Intel Tofino
ASIC~\cite{tofino}. The control plane is implemented in Python and
interacts with the switch data plane through Thrift APIs~\cite{thrift} using Intel's P4 
SDE 9.10.0. We implement the tables as described in $\S$\ref{subsec:sdf} using 1.3MB SRAM in the switch. 
The GC states of the replica and destination table use registers, 
such that they can be updated in the data plane, consuming a total of 128KB of
stateful memory.

Since we do not have access to a real data center, we emulate datacenter network traffic in our
cluster using traces and released network traffic distributions~\cite{mogul:hotos2015,iqbal:ton2022,popescu:tma2018}. 
%(BEN) add clear description of network evaluation
%    Two distributions -- \mbox{\cite{mogul:hotos2015,iqbal:ton2022}}-- report round trip times,
%while one --\mbox{\cite{popescu:tma2018}}-- reports one-way delay. 
The traces include delays between VMs in cloud data centers. 
%We use \mbox{\cite{popescu:tma2018}} as a baseline since it reports latencies over time. 
To simulate the variations of network latency, we scale the trace in \mbox{\cite{popescu:tma2018}} following the latency patterns and distributions in
\mbox{\cite{mogul:hotos2015,iqbal:ton2022}}. The latency is associated with each request and stored in the LAT field 
when the packet is generated (see Figure~\mbox{\ref{fig:packetformat}}). 
When the request traverses the switch, we add the per-hop latency as described in
$\S$\mbox{\ref{subsec:ioschedule}}. The end-to-end latency is computed by adding the time spent at the
storage server and the final LAT value in the return packet. 
%(BEN) this was here before
%We emulate these traffic behaviors by generating requests with embedded \emph{LAT} values 
%following the observed distributions. 

\noindent
\textbf{Storage implementation.}
We build the SDF (SSD virtualization) stack on top of programmable SSDs. 
%They allow issuing reading/writing/erasing commands. 
By default we implement a greedy, threshold-based GC. We specify
the GC thresholds used in each experiment in $\S$\ref{sec:eval}. 
%For background GC, we predict every $30$ seconds when the SDF checks whether GC should be triggered. The prediction model uses minimal
%resources, consuming 16KB memory and requiring $0.7$ milliseconds for inference, and $11.3$ milliseconds for training.
%%These overheads are incurred once every five seconds (for up to $2048$ vSSDs per server). 
%We use $5$ minutes to warmup the predictor. %before issuing requests for background GC.
%We use PyTorch v1.9.0~\cite{pytorch} to implement the predictors. They have a
%hidden LSTM layer of $24$ nodes with fully connected input/output layers. We apply Softmax to the
%output, and use %\textit{adam}~\cite{adam:arxiv} and 
%mean squared error as the loss function. 
In our testbed, we use one server as clients, and others to host vSSDs.
%For each workload we evaluate, we modify the backend to track the storage operations they issue and
%forward them to the client, to send the vSSD.

\noindent
\textbf{Emulation.}
Since we only have one type of programmable SSD, we build an SSD emulator using Python to test
\pname{} against different SSD device performance (see $\S$\ref{subsubsec:eval_lat_sens}). 
%(BEN) adding some detail on emulation
We validate the emulator with our programmable SSD.  
%and change the hardware parameters to match the performance of Intel DC and Optane.}
For these experiments, we use the same implementation, but issue requests to the emulated SSDs.  
%instead of the programmable SSD.

\noindent
\textbf{Others.} Similar to modern storage
systems~\mbox{\cite{ghemawat:sosp2003,www-hadoop-arch}}, \mbox{\pname{}} leverages heartbeats to
detect failures. On
link failure, it redirects requests to replicas in the rack. 
On server failure,
\mbox{\pname{}} replicates the replicas to other servers and updates their
switches.
%(BEN) add the server recovery process
Upon data recovery, it updates stale data from replica vSSDs before
serving requests~\mbox{\cite{hermes:asplos2020}}. 
On switch failure, \mbox{\pname{}} relies on replicas in another rack to serve requests.
The ToR switch is repopulated on switch recovery. \mbox{\pname{}} focuses on 
storage management of a rack. As future work, we wish to extend it to multiple racks by modifying 
%Algorithm~\mbox{\ref{algo:sdn}} to forward GC packets to keep GC states consistent among switches.
Algorithm~\mbox{\ref{algo:sdn}} to keep GC states consistent among switches.

%\vspace{-1ex}

\section{Evaluation}
\label{sec:eval}
%OUTLINE
Our evaluation shows that:
(1) \pname{} reduces the tail latency of I/O requests by up to 5.8$\times$ for data-center
applications with network-storage coordination (\S\ref{subsec:eval_perf} and
\S\ref{subsec:eval_various_workloads});
(2) \pname{} works with various storage and network scheduling policies
(\S\ref{subsubsec:eval_storage_schedulers} and \S\ref{subsubsec:eval_net_schedulers});
(3) \pname{} benefits various SSD devices and network latency distributions
(\S\ref{subsubsec:eval_lat_sens}); 
%(4) \pname{} improves isolation for software-isolated vSSDs (\S\ref{subsubsec:eval_gc_iso});
and (4) \pname{} extends the lifetime of a rack of SSDs (\S\ref{subsec:eval_wear}).
\begin{table}[t]
    \centering
    \footnotesize
    \caption{Workloads used in our evaluation.}
    \vspace{-3.5ex}
    \begin{tabular}{|l|l|l|}
        \hline
        {\bf Workload} & {\bf Description} & {\bf Write\%} \\\hline
        YCSB~\cite{www-ycsb} & Cloud data serving queries. & 0-100\% \\\hline
        TPC-H~\cite{www-tpch} & Business-oriented ad-hoc queries. & 2.27\% \\\hline
        Seats~\cite{www-seat} & Airline ticketing system queries. & 10.34\% \\\hline
        AuctionMark~\cite{www-auction} & Activity queries in an auction site. & 53.76\% \\\hline
        TPC-C~\cite{www-tpcc} & Online transaction queries. & 59.95\% \\\hline
        Twitter~\cite{www-twitter-bench} & Micro-blogging website queries.& 97.86\% \\\hline
    \end{tabular}
    \vspace{-3.5ex}
    \label{tab:real_workloads}
\end{table}

\vspace{-0.5ex}
\subsection{Experimental Setup}
\label{subsec:exp_setup}

%OUTLINE
To examine \pname{}'s performance under different workload patterns, we use YCSB with
different read/write ratios~\cite{www-ycsb} and various workloads from BenchBase~\cite{benchbase}.
These represent common data center applications sensitive to network and storage performance
(Table~\ref{tab:real_workloads}). All workloads run on hardware-isolated vSSDs. 
%\hl{The benefits of \mbox{\pname{}} apply similarly for hardware-isolated and software-isolated vSSDs}. 
The datasets range in 50-100GB, so we allocate vSSD capacity accordingly (64-128GB). We set
\textit{soft\_threshold} to 35\%, and set \textit{gc\_threshold} to 25\%. 
Before each experiment, we run a subset of the workloads to trigger GC and consume 50\% of the free blocks.
%we run a subset of the workloads to warm up the SSD to trigger GC.
%Unless otherwise specified, t
%After warmup, 50\% of the free blocks are consumed. 
\pname{} uses Linux's Kyber scheduler~\mbox{\cite{kyber}} by default, as it performs the best across various settings (see
\S\mbox{\ref{subsubsec:eval_storage_schedulers}}). 
Kyber uses 750$\mu$s for reads and 3 millisecs for writes as target 95-th percentile (P95) latencies.
When enabling coordinated I/O, we use 1.75 millisecs and 4 millisecs to account for P95 network delay.
We use the default priority-based isolation in the switch. 

To show the performance benefits of network-storage co-design, we compare \pname{} with state-of-the-art 
software-defined storage architecture designs at datacenter scale. 
%Specifically, we evaluate the following designs:

\textbf{VDC}: Virtual datacenter (VDC)~\mbox{\cite{virtualdc:osdi2014}} enables end-to-end 
isolation between multiple tenants sharing the same physical network and storage. It implements a logically centralized
controller that allocates resources to each tenant's VDC as well as each tenant's I/O 
flows~\cite{eno:sosp2013}. We run the controller on a separate server updating flow demand and allocations. VDC enforces
end-to-end isolation for each flow with multi-resource token bucket rate limiting.

\textbf{\pname{ (Software)}}: Although \pname{} is developed with a programmable switch and SSDs, its core ideas can  
be implemented in the software stack. To evaluate this, we extend VDC by adding software-based coordinated I/O scheduling 
and GC.  We make the VDC controller GC-aware by tracking the GC state of vSSDs, and implementing the coordinated GC ($\S$\mbox{\ref{subsec:gc}}) 
in software. When the controller grants the vSSD's request to perform
GC, it also returns the location of a replica not performing GC. Therefore, storage servers can 
redirect requests when the vSSD is performing GC.

\subsection{End-to-End Performance Benefits}
\label{subsec:eval_perf}

\begin{figure}[t]
    \centering
    \begin{subfigure}{\linewidth}
        \centering
        \includegraphics[width=0.9\linewidth]{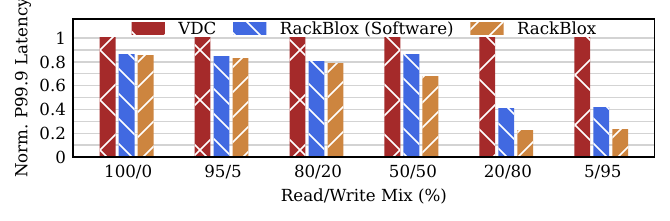}
        \vspace{-2.5ex}
        \caption{P99.9 read latency (excluding write-only).}
        \label{fig:eval_p999_tail_lat_read_less_gc}
    \end{subfigure}
    \vfill
    \begin{subfigure}{\linewidth}
        \centering
        \includegraphics[width=0.9\linewidth]{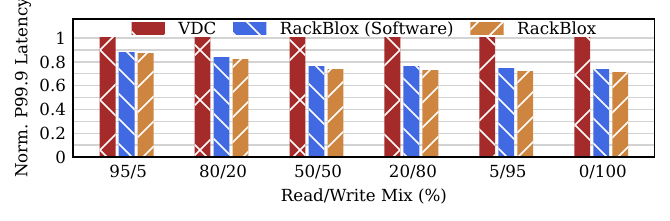}
        \vspace{-2.5ex}
        \caption{P99.9 write latency (excluding read-only).}
        \label{fig:eval_p999_tail_lat_write_less_gc}
        \vspace{-2ex}
    \end{subfigure}
    \caption{\pname{}'s benefits for P99.9 end-to-end latency.}% against state-of-the-art .}
    \label{fig:eval_p999_tail_lat_less_gc}
	\vspace{-3.5ex}
\end{figure}

%FIGURE 9a:
%VDC:          2805,2922,3156,3885,11948,12358
%VDC-Enhanced: 2755,2888,3071,3820,4854,5184
%RackBlox:     2400,2418,2476,2620,2716,2822
%FIGURE 9b:
%VDC:          3094,3095,3931,3990,4094,4262
%VDC-Enhanced: 3094,3095,3938,3992,4098,4308
%RackBlox:     2687,2541,2888,2890,2946,3010

%(BEN) adding the raw latencies for the relative speedup comparisons
To evaluate the end-to-end performance of \mbox{\pname{}}, we run YCSB
benchmarks with the zipfian request distribution, and vary the write ratio from 0\% (read-only) to 100\% (write-only).
%We show the reduction in end-to-end tail latency when the SSD is 50\% full in
%Figure~\mbox{\ref{fig:eval_p999_tail_lat_less_gc}}.
With network-storage co-design, \mbox{\pname{}} improves the 99.9-th percentile (P99.9) read latency
by up to 4.4$\times$ (12.4 millisecs vs. 2.8 millisecs), the P99.9 write latency by up to
1.4$\times$ (4.3 millisecs vs. 3.0 millisecs), as shown in 
%(BEN) Adding this here, the reason to not adjust the figure is that we have a range of 2-12 ms
%latencies and the issue is that its very hard to see any speedup at low read rates with that y-axis
Figure~\ref{fig:eval_p999_tail_lat_less_gc}. 
%\hl{We show normalized latencies to highlight the benefits across the different read/write distributions.}
We show the detailed results in Figure~\mbox{\ref{fig:eval_lat_cdf_breakdown}}.

%Figure~\mbox{\ref{fig:eval_p999_tail_lat_read_less_gc}} shows the reduction in end-to-end read tail
%latency. 
Although VDC ensures flow isolation across network and storage stack in software, %However, while it can guarantee performance within each stack, 
it performs worse than \pname{} (Figure~\ref{fig:eval_p999_tail_lat_read_less_gc}), 
due to the lack of the coordination between the network and storage stack.
\pname{ (Software)} enables GC redirection in software, and reduces the overhead when requests
are blocked by GC. For read-heavy workloads, the performance of \pname{ (Software)} is similar to that of \pname{}, 
as they both conduct the coordinated I/O scheduling. 
%we see no benefit as it cannot consider the network latency for coordinated I/O scheduling. 
However, for write-heavy workloads, which cause more intensive
GC, \pname{ (Software)} can improve VDC's performance by to $2.4\times$ (12.4 millisecs vs. 5.2
millisecs). 
However, since \pname{ (Software)} incurs additional network overhead, it remains suboptimal.
%, compared to \pname{} developed within the programmable switch. 
\pname{} outperforms VDC and \pname{ (Software)} by enabling 
coordinated I/O scheduling and coordinated GC in the network by $4.4\times$ and
$1.84\times$ (5.2 millisecs vs. 2.8 millisecs) respectively. 

\begin{figure}[t]
    \centering
    \begin{subfigure}{\linewidth}
        \centering
        \includegraphics[width=0.9\linewidth]{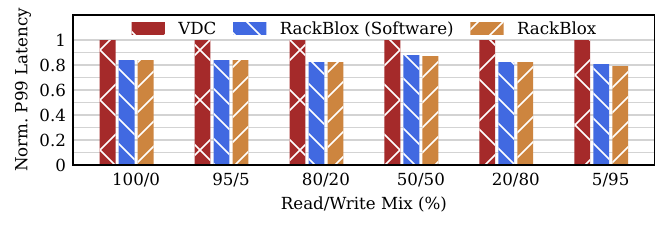}
        \vspace{-2ex}
        \caption{P99 read latency (excluding write-only).}
        \label{fig:eval_p99_tail_lat_read_less_gc}
    \end{subfigure}
    \vfill
    \begin{subfigure}{\linewidth}
        \centering
        \includegraphics[width=0.9\linewidth]{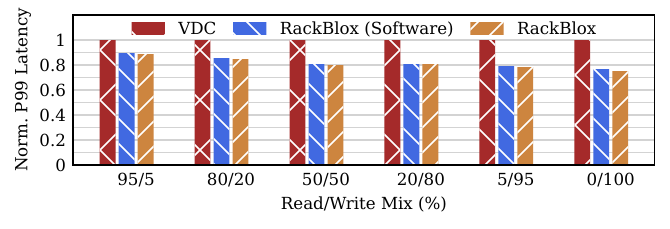}
        \vspace{-2ex}
        \caption{P99 write latency (excluding read-only).}
        \label{fig:eval_p99_tail_lat_write_less_gc}
        \vspace{-2ex}
    \end{subfigure}
    \caption{\pname{}'s benefits for P99 end-to-end latency.}% against state-of-the-art .}
    \label{fig:eval_p99_tail_lat_less_gc}
	\vspace{-2ex}
\end{figure}

Writes are hardly affected by GC because of the write cache in storage servers, as shown in
Figure~\mbox{\ref{fig:eval_p999_tail_lat_write_less_gc}}. Thus, \pname{} and \pname{ (Software)} 
have similar performance, as the coordinated I/O scheduling improves the 99.9th percentile write tail 
latency by up to $1.4\times$.
We show \mbox{\pname{}}'s benefit for the 99th percentile (P99) latency in
Figure~\mbox{\ref{fig:eval_p99_tail_lat_less_gc}}.
The read latency is improved by up to 2.1$\times$ (5.3 millisecs vs.
2.6 millisecs) and the write latency is improved by up to 1.3$\times$ (3.7 millisecs vs. 2.8 millisecs). 
This demonstrates that \mbox{\pname{}} can achieve benefit at lower tails as well. 

\begin{figure}[t]
    \centering
    \begin{subfigure}{\linewidth}
        \centering
        \includegraphics[width=0.9\linewidth]{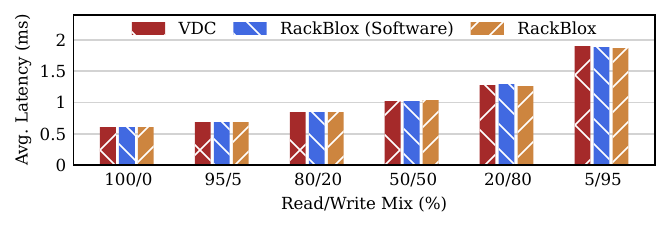}
        \vspace{-2ex}
        \caption{Average read latency (excluding write-only).}
        \label{fig:eval_avg_lat_read_less_gc}
    \end{subfigure}
    \vfill
    \begin{subfigure}{\linewidth}
       \centering
        \includegraphics[width=0.9\linewidth]{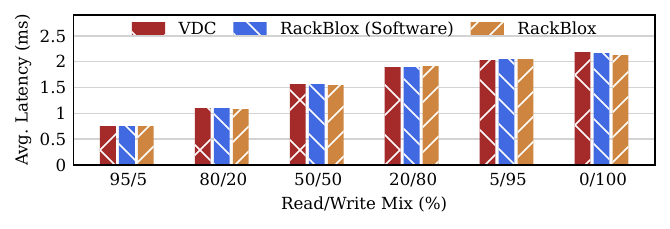}
        \vspace{-2ex}
        \caption{Average write latency (excluding read-only).}
        \label{fig:eval_avg_lat_write_less_gc}
        \vspace{-2ex}
    \end{subfigure}
	\caption{Comparing \pname{}'s average end-to-end latency against VDC and \pname{ (Software)}.}
    \label{fig:eval_avg_lat_less_gc}
	\vspace{-2ex}
\end{figure}

\mbox{\pname{}} does not negatively affect the average latency, as shown in Figure~\mbox{\ref{fig:eval_avg_lat_less_gc}}. 
As we increase the write ratio in the workloads, the average latency of reads/writes is gradually increased, due to the read/write 
interference, and the write latency is longer than read latency in the storage stack.

\begin{figure}[t]
    \centering
    \includegraphics[width=0.9\linewidth]{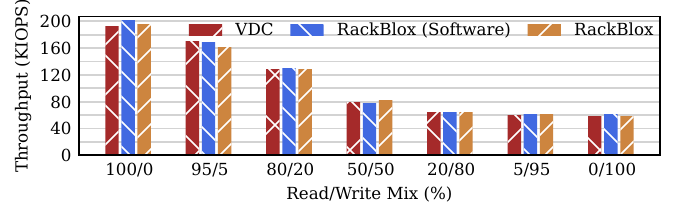}
    \vspace{-3.5ex}
    \caption{\pname{}'s impact on throughput.}
    \label{fig:eval_tput_less_gc}
    \vspace{-2ex}
\end{figure}

We show the average throughput of YCSB benchmarks in Figure~\mbox{\ref{fig:eval_tput_less_gc}}. 
\mbox{\pname{}} does not negatively affect throughput, as 
\mbox{\pname{}} targets improved tail latency. 
Similar to the average latency, higher write rates lead to lower IOPS since writes have higher device latency. 

\subsection{End-to-End Performance of Various Workloads}
\label{subsec:eval_various_workloads}

\begin{figure}[t]
    \centering
    \begin{subfigure}{\linewidth}
        \centering
        \includegraphics[width=0.9\linewidth]{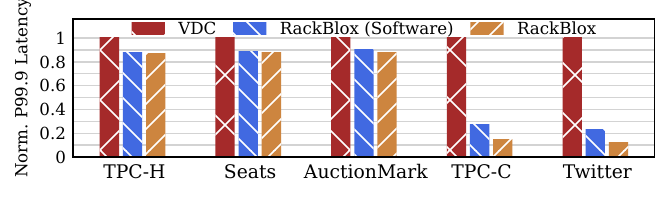}
        \vspace{-2ex}
        \caption{P99.9 read latency (excluding write-only).}
        \label{fig:eval_real_workload_tail_lat_read}
    \end{subfigure}
    \vfill
    \begin{subfigure}{\linewidth}
        \centering
        \includegraphics[width=0.9\linewidth]{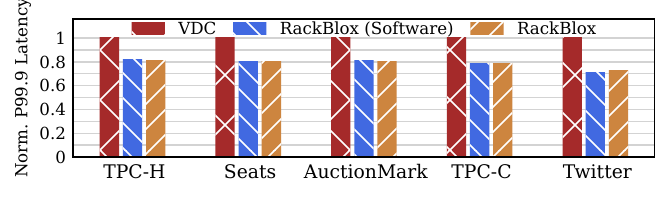}
        \vspace{-2ex}
        \caption{P99.9 write latency (excluding read-only).}
        \label{fig:eval_real_workload_tail_lat_write}
        \vspace{-2ex}
    \end{subfigure}
	\caption{Comparing \pname{}'s P99.9 end-to-end latency against VDC and \pname{ (Software)} for various workloads.}
    \label{fig:eval_real_workload_tail_lat}
	\vspace{-3ex}
\end{figure}

%FIGURE 11a:
%VDC:          2842,3158,3100,19800,23818
%VDC-Enhanced: 2842,3158,3100,4442,5126
%RackBlox:     2462,2770,2716,2872,2996
%FIGURE 11b:
%VDC:          4480,4140,4254,4296,4492
%VDC-Enhanced: 4480,4140,4254,4292,4492
%RackBlox:     3132,3100,3124,3136,3162

\begin{figure}[t]
    \centering
    \includegraphics[width=0.9\linewidth]{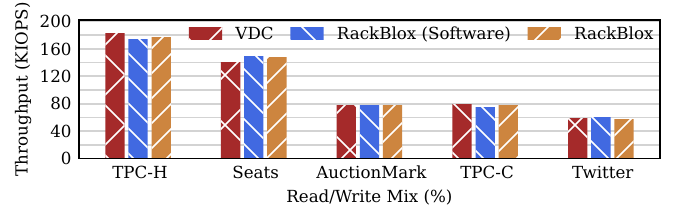}
    \vspace{-3ex}
    \caption{Throughput for various workloads.}
    \label{fig:eval_real_workload_tput}
    \vspace{-3ex}
\end{figure}

We further evaluate \mbox{\pname{}} on various workloads (see Table~\ref{tab:real_workloads}),
following the setup described in \S\mbox{\ref{subsec:exp_setup}}. 
Figure~\mbox{\ref{fig:eval_real_workload_tail_lat_read}} shows that \mbox{\pname{}} improves the
P99.9 read latency of various workloads by up to 7.9$\times$ (23.8 millisecs vs. 3.0 millisecs), in comparison with VDC.  
For P99 read tail latency, \mbox{\pname{}} achieves up to 2.9$\times$ improvement (8.1 millisecs vs. 2.8 millisecs). 
Compared to the YCSB experiments, we observe similar correlation between 
write ratio and read tail latency improvement in various workloads.
For read-intensive workloads like TPC-H, \mbox{\pname{}} and \pname{ (Software)} improve mainly via coordinated I/O scheduling. 
%And we obtain similar performance benefits with \pname{ (Software)}. 
For write-intensive ones like Twitter, \pname{} improves performance mainly by alleviating
GC interference. % its software-based implementation, 
AuctionMark has less benefit than YCSB with 50\% writes, although it has
slightly higher write ratio. This is because AuctionMark has a different I/O request pattern (e.g., a long sequence of writes followed by a sequence of reads, 
rather than mixed reads and writes in YCSB), it has fewer I/O requests affected by the GC. 
\pname{ (Software)} performs worse than \pname{} due to the 
additional networking overhead for coordinated GC. 
The end-to-end write tail latency, shown in
Figure~\mbox{\ref{fig:eval_real_workload_tail_lat_write}}, demonstrates a similar trend and improvement to YCSB.
For throughput (see Figure~\mbox{\ref{fig:eval_real_workload_tput}}) and average read/write latency, 
we observe the similar trend as 
YCSB benchmarks (see $\S$\ref{subsec:eval_perf}).

\subsection{Performance Benefit Breakdown in \pname{}}
\label{subsec:eval_perf_analysis}

\begin{figure}[t]
    \centering
    \includegraphics[width=1.\linewidth]{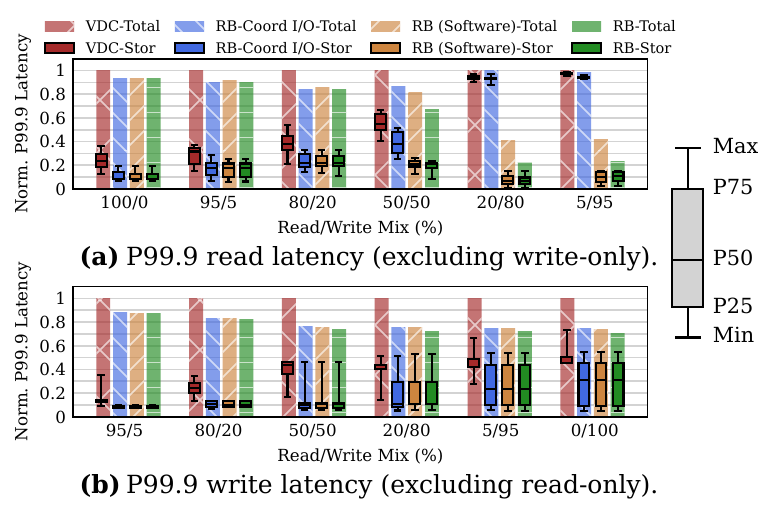}
   \vspace{-5ex}
    \caption{P99.9 latency breakdown. RB is \mbox{\pname{}}. Stor is the storage latency and Total is the end-to-end latency.}
    \label{fig:eval_p999_tail_lat_breakdown}
	\vspace{-3ex}
\end{figure}
%\begin{figure}[t]
%    \centering
%    \begin{subfigure}{\linewidth}
%        \centering
%        %\includegraphics[width=0.9\linewidth]{./Figures/eval_p999_tail_lat_read_breakdown2.pdf}
%        \includegraphics[width=0.9\linewidth]{./Figures/eval_boxplot_read_norm.pdf}
%        \vspace{-2.5ex}
%        \caption{P99.9 read latency (excluding write-only).}
%        \label{fig:eval_p999_tail_lat_read_breakdown}
%    \end{subfigure}
%    \vfill
%    \begin{subfigure}{\linewidth}
%        \centering
%        %\includegraphics[width=0.9\linewidth]{./Figures/eval_p999_tail_lat_write_breakdown.pdf}
%        \includegraphics[width=0.9\linewidth]{./Figures/eval_boxplot_write_norm.pdf}
%        \vspace{-2.5ex}
%        \caption{P99.9 write latency (excluding read-only).}
%        \label{fig:eval_p999_tail_lat_write_breakdown}
%        \vspace{-2.5ex}
%    \end{subfigure}
%    \caption{\hlgreen{P99.9 latency breakdown. RB is \mbox{\pname{}}. Stor is the storage latency
%    and Total is the end-to-end latency.}}
%    \label{fig:eval_p999_tail_lat_breakdown}
%	\vspace{-3ex}
%\end{figure}

%FIGURE 13:
%VDC:          2583,2700,2954,3885,11972,12358
%RB-Coord IO:  2400,2418,2476,3170,11972,12123
%RB (Software):2411,2468,2542,3150,4914,5184
%RackBlox:     2400,2418,2476,2620,2716,2822

\begin{figure}[t]
    \centering
    \includegraphics[width=0.95\linewidth]{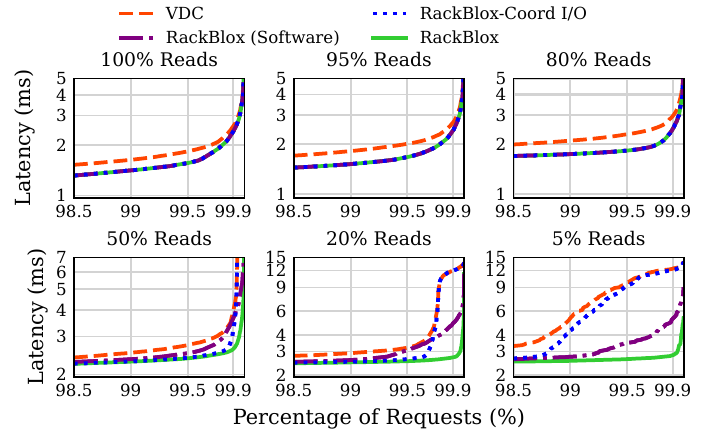}
    \vspace{-3.5ex}
    \caption{Cumulative distribution of read latency.}
    \label{fig:eval_lat_cdf_breakdown}
    \vspace{-3ex}
\end{figure}

%\begin{figure}[t]
%    \centering
%    \includegraphics[width=0.9\linewidth]{./Figures/eval_sw_gc_read.pdf}
%    \vspace{-3ex}
%    \caption{\hl{Comparing P99.9 read latency for \mbox{\pname{}} against software-only when the SSD
%    is 50\% full (excluding write-only)}.}
%    \label{fig:eval_sw_gc_read}
%    \vspace{-2ex}
%\end{figure}

%We now analyze the source of the performance improvements in \mbox{\pname{}} 
%using the YCSB benchmarks. To facilitate our analysis, we 
To break down the performance improvements in \mbox{\pname{}}, we
evaluate \textbf{\mbox{\pname{}}-Coord I/O}, in which we enable coordinated I/O scheduling between the network and storage stack, 
but disable coordinated GC in \pname{}.

We show the comparison in Figure~\mbox{\ref{fig:eval_p999_tail_lat_breakdown}} with latency breakdowns. 
%While VDC can provide instance-level isolation through the network and storage stack, 
%the inability to coordinate between network and storage results in I/O requests 
%with high network latency being delayed in the storage stack. Therefore, it has the worst
%performance.
%(BEN) add the analysis of breakdowns
%\hl{The storage latency has lower contribution at low write ratios and increases 
%until GC overhead dominates the tail latency at high write ratios.}
%We observe that \mbox{\pname{}}-Coord I/O outperforms
%\pname{ (Software)} for read-intensive workloads (over 20\% reads). 
By coordinating I/O between network and storage, \mbox{\pname{}}-Coord I/O 
%(BEN) discussed later anyways
%and \pname{ (Software)} 
reduces the P99.9 read latency by up to 1.1-1.23$\times$ (3.9 millisecs vs. 3.1 millisecs) and write
latency by 1.1-1.4$\times$ compared to VDC.  %, as shown in Figure\mbox{\ref{fig:eval_p999_tail_lat_breakdown}}.}
%(shown in Figure~\mbox{\ref{fig:eval_p999_tail_lat_write_breakdown}})}. 
With increased write ratios, \mbox{\pname{}}-Coord I/O brings more benefits for the tail latency, 
because the potential delay of a request in the storage queues increases, as writes have higher device
latency than reads. 
Thus, prioritizing requests in the storage queue leads to more obvious effects on
the end-to-end latency and coordinated I/O scheduling provides greater speedup.
%(BEN) add analysis of the breakdowns 
%\hl{This pattern is shown in the breakdown for VDC and \mbox{\pname{}}-Coord
%I/O where \mbox{\pname{}-Coord I/O} reduces the storage latency by prioritizing requests that have
%high network latency to achieve greater benefit.}

However, for write-dominant workloads (e.g., more than 50\%), the read tail latency improvement of coordinated I/O scheduling 
diminishes no matter how we schedule as shown in
Figure~\mbox{\ref{fig:eval_p999_tail_lat_breakdown}a}, because intensive writes incur high GC overhead. 
%As shown in Figure \mbox{\ref{fig:eval_lat_cdf_breakdown}}, with more than 80\% writes, at least
%0.25\% of reads are GC blocked. 
%Since GC overhead dominates end-to-end latency of these requests, coordinated I/O scheduling can 
%do little to improve their latencies. 
With high write ratios, the coordinated I/O scheduling brings more benefits to the tail latency of writes than  
that of reads, since the write cache helps alleviate the GC overhead.  
%As we increase the write ratio, the writes dominate the queue delay, 
%and the benefit of coordinated I/O scheduling is not further increased.} 
As we further increase the write ratio (i.e., above 50\%), the tail latency of both VDC and RackBlox-Coord I/O is increased, due to 
the increased storage queue delay (see the storage latency distribution in
Figure~\mbox{\ref{fig:eval_p999_tail_lat_breakdown}b}). 
Compared to VDC, the normalized tail latency reduction of coordinated I/O scheduling is almost the 
same as shown in Figure~\mbox{\ref{fig:eval_p999_tail_lat_breakdown}b}.

The coordinated GC mechanism in \pname{} will 
further improve the read tail latency (by up to 4.3$\times$), as shown in
Figure~\ref{fig:eval_p999_tail_lat_breakdown}a 
and Figure~\ref{fig:eval_lat_cdf_breakdown}.
Both \pname{} and \pname{ (Software)} implement coordinated GC, but \pname{} provides more speedup with the programmable switch, as it alleviates the
unnecessary networking round-trip delays. 
The coordinated GC does not benefit writes, as we need to issue writes to all replicas for data consistency (as discussed in
$\S$\mbox{\ref{subsubsec:hw_gc}}).

\vspace{-1.0ex}
\subsection{Sensitivity Analysis}
\label{eval:sens}
We demonstrate that \mbox{\pname{}} retains the flexibility and modularity of the original
SDN/SDF design by evaluating different scheduling policies and system configurations. 

\vspace{-1.5ex}
\subsubsection{Varying storage I/O scheduling policies.}
\label{subsubsec:eval_storage_schedulers}
\begin{figure}[t]
    \centering
    \begin{subfigure}{\linewidth}
        \centering
        \includegraphics[width=0.88\linewidth]{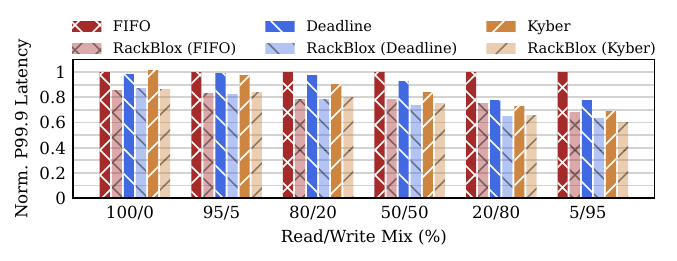}
        \vspace{-2.5ex}
        \caption{P99.9 read latency for YCSB (excluding write-only).}
        \label{fig:eval_p999_tail_lat_read_sched}
    \end{subfigure}
    \vfill
    \begin{subfigure}{\linewidth}
        \centering
        \includegraphics[width=0.88\linewidth]{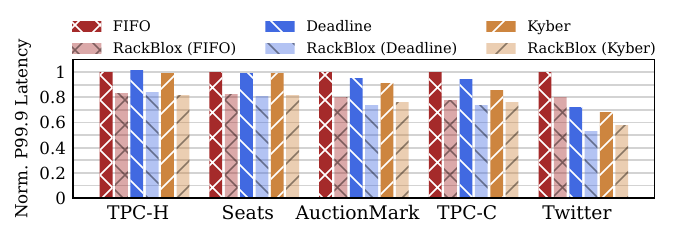}
        \vspace{-2.5ex}
        \caption{P99.9 read latency for various workloads.}
        \label{fig:eval_real_workload_sched_tail_lat_read}
        \vspace{-3.0ex}
    \end{subfigure}
    \caption{Comparing \pname{}'s P99.9 end-to-end latency with different storage I/O schedulers.}
    \label{fig:eval_sched_read_tail_lat}
	\vspace{-3.0ex}
\end{figure}

We now examine the benefit of the coordinated I/O scheduling  
under different storage I/O scheduling policies. In particular, 
we implement Linux's storage schedulers for SDF:
no-op (\textbf{FIFO}), \textbf{Deadline}, and \textbf{Kyber}~\mbox{\cite{kyber}}. 
No-op is the default on NVMe devices, while both Kyber and Deadline 
target predictable latency. Deadline splits requests into read and write queues, and 
prioritizes requests when they reach their respective deadlines. 
Kyber also splits requests into read and write queues, and
throttles each queue to meet the latency targets. To enable coordinated
I/O scheduling, \pname{} reorders requests in each queue using network latencies.
We use 0.5 millisecs and 1.75 millisecs as deadlines for reads and writes in Deadline and 1.5
millisecs and 2.75 millisecs in \pname{} (Deadline). 
We use 0.75 millisecs and 3 millisecs for reads and writes in Kyber and 1.75 millisecs and 4 millisecs for
\pname{} (Kyber). 
%(BEN) Make more clear why there are different parameters and how they were selected.
\mbox{\pname{}} (Deadline) and \mbox{\pname{}} (Kyber) use increased parameter values, as \mbox{\pname{}} 
incorporates the network latency in its coordinated I/O scheduling, based on the distribution of network latencies in data
centers~\mbox{\cite{popescu:tma2018}}.  
%While the performance was not sensitive to these
%parameter, the selected parameter values had the best performance in our experiments.}
%For Kyber-Coord and Deadline-Coord, we add 1 millisec to their targets to account for the
%network. These values help prioritize reads while ensuring writes are flushed with reasonable latency.

We show the results in Figure~\mbox{\ref{fig:eval_sched_read_tail_lat}}. As expected,  
coordinated I/O scheduling always outperforms its baseline scheduler. 
%In fact, any coordinated I/O scheduler almost always outperforms all baseline schedulers. 
\pname{ (FIFO)} achieves the greatest speedup over its baseline scheduler (1.5$\times$). 
\pname { (Kyber)} and \pname{ (Deadline)} have fewer opportunities to reorder
requests when splitting reads and writes into separate queues, but still benefit from coordination
(1.24$\times$ and 1.36$\times$ respectively).
%Since kyber and deadline split requests into
%reads/writes, there are fewer opportunities to reorder within these queues compared to FIFO-Coord.}

%FIGURE 15a:
%FIFO:           2805,2922,3156,3377,3636,4190
%FIFO-Coord:     2385,2419,2476,2644,2732,2830
%Deadline:       2755,2888,3071,3121,2826,3254
%Deadline-Coord: 2435,2399,2474,2494,2368,2636
%Kyber:          2845,2842,2851,2826,2638,2890
%Kyber-Coord:    2410,2448,2528,2540,2382,2502
%FIGURE 15b:
%FIFO:           2974,3056,3332,3418,2334
%FIFO-Coord:     2474,2514,2660,2648,1870
%Deadline:       3006,3032,3166,3226,1684
%Deadline-Coord: 2494,2464,2458,2522,1236
%Kyber:          2936,3010,3018,2910,1584
%Kyber-Coord:    2430,2494,2540,2586,1336

\subsubsection{Varying network scheduling policies.}
\label{subsubsec:eval_net_schedulers}
\begin{figure}[t]
    \centering
    \includegraphics[width=0.9\linewidth]{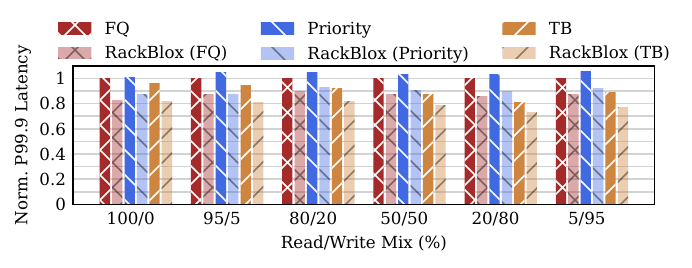}
    \vspace{-2.5ex}
    \caption{Comparing P99.9 read latency for \pname{} with different network scheduling policies.}
    \label{fig:eval_net}
    \vspace{-2ex}
\end{figure}
We now evaluate the performance of \mbox{\pname{}} under different network scheduling policies in the switch.
Besides the Token Bucket rate limiting (\textbf{TB}) policy that ensures isolation between flows
(similar to VDC), we examine the fair queuing (\textbf{FQ}) and
priority based network scheduling (\textbf{Priority}) policies. 
For FQ, we have four client servers competing for one
storage server with each receiving a fair share of the network bandwidth. In Priority, we periodically
create higher priority traffic using~\mbox{\cite{socialnetwork:sigcomm2015}}, which delays
lower-priority requests.
%We use INT to add the packet delays in the switch.

We show the results in Figure~\mbox{\ref{fig:eval_net}}. Coordinated I/O scheduling 
can benefit all the underlying network schedulers. FQ and Priority result
in higher latency as requests are delayed in the network. This provides increased
opportunities for reordering, which allows \pname{} to achieve up to 1.21$\times$ and 1.15$\times$ performance 
improvement on average, respectively. 
%We observe similar performance improvements for all three network schedulers see similar improvements 
%(up to 1.21$\times$ and 1.15$\times$ on average).

%FIGURE 16:
%Fair Queuing:          2964,3023,3100,3232,3251,3248
%Fair Queuing-Coord:    2444,2631,2782,2830,2782,2824
%Priority Based:        2979,3175,3244,3332,3340,3439
%Priority Based-Coord:  2591,2640,2880,2910,2922,2994
%Token-Bucket:          2845,2842,2851,2826,2638,2890
%Token-Bucket-Coord:    2410,2448,2528,2540,2382,2502

\begin{figure}[t]
    \centering
    \includegraphics[width=0.9\linewidth]{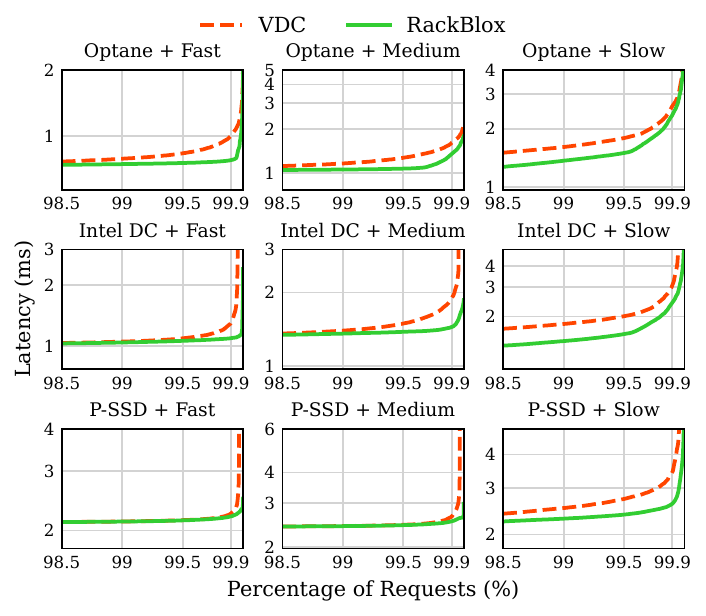}
    \vspace{-2.5ex}
    \caption{End-to-end read latency distributions of \pname{} on YCSB-A with varying SSDs and network latencies.}
    \label{fig:eval_sens_cdf}
    \vspace{-2.5ex}
\end{figure}
\subsubsection{Varying the Network/Storage Latency}
\label{subsubsec:eval_lat_sens}

The coordinated I/O scheduling works by hiding higher network latency with lower storage latency, or vice versa.
%, assuming the average network and storage latencies are similar. 
Therefore, if the network latency overwhelms storage latency or vice versa, \pname{} helps less to improve the end-to-end latency.

%(BEN) adding emulated here for clarity
We analyze the sensitivity of \pname{} by evaluating the YCSB benchmarks with emulated devices of different
latencies (see $\S$\mbox{\ref{subsec:implt}}). We evaluate three SSDs from fastest to slowest:
\textbf{Optane}~\cite{intel_optane_900p_ssd}, \textbf{Intel DC}~\cite{intel_dc_ssd}, and
\textbf{P-SSD} (programmable SSD)~\cite{bjorling:fast2017}. We evaluate networks 
with \textbf{Fast}~\cite{popescu:tma2018}, \textbf{Medium}~\cite{mogul:hotos2015}, and
\textbf{Slow}~\cite{iqbal:ton2022} latencies (see $\S$\mbox{\ref{subsec:implt}}). The resulting
end-to-end latencies of YCSB-A (50\% reads) are shown in Figure~\ref{fig:eval_sens_cdf}.
\begin{figure}[t]
    \centering
    \includegraphics[width=0.9\linewidth]{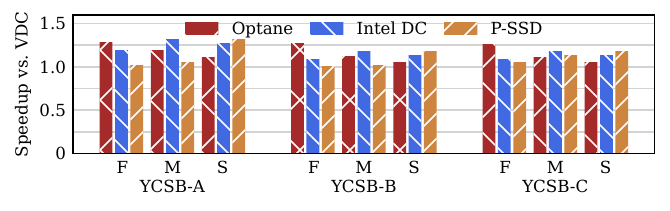}
    \vspace{-2.5ex}
    \caption{P99.9 read latency improvements of \pname{} with different storage and network latencies.}
    \label{fig:eval_sens}
    \vspace{-2.5ex}
\end{figure}

% hides the variation of storage latency with that of network, or vice versa. Thus, if one variation significantly outweighs another, the dominating variation is not hidden and coordination benefits P99.9 latency less. Thus, we observe that when the SSD or the network is much slower than the other, \pname{} is less effective, as represented by the bottom-left and top-right corner of Figure \ref{fig:eval_sens_cdf}. In contrast, when storage and network have matching performances, \pname{} brings the most benefits. 

\noindent
\textbf{Varying the SSD performance.}
For \pname{}, the marginal benefit on end-to-end tail latency by upgrading SSD is low when the SSD
already outperforms the network. For example, %in Figure~\ref{fig:eval_sens_cdf}, 
upgrading the
SSD from Intel DC to Optane under Slow network brings little benefit to the P99.9 latency.
%since the bottleneck is the slow network. 
Thus, the performance improvement of \pname{} over VDC is also low. 
In contrast, the benefits of upgrading SSD are more obvious when the network outperforms SSDs, 
which brings overall performance benefit for \pname{}.

\label{subsubsec:eval_sw_iso}
\begin{figure}[t]
    \centering
    \includegraphics[width=0.70\linewidth]{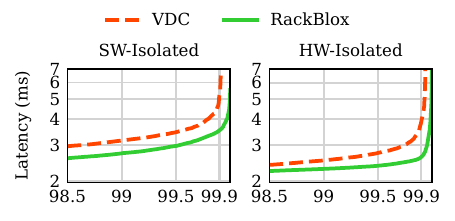}
    \vspace{-3ex}
    \caption{Read tail latency with different isolation.}
    \label{fig:eval_sw_hw_cdf}
    \vspace{-3ex}
\end{figure}

\noindent
\textbf{Varying the networking performance.}
Similar conclusions are drawn as we vary network latencies. For example, upgrading the network from Slow to Fast with the slowest P-SSD hardly improves read tail latency in \pname{}, because SSD latency dominates the end-to-end latency. In contrast, by upgrading the network with the fastest Optane SSD, we significantly improve the read tail latency in \pname{}.

Our findings are consistent across various YCSB benchmarks, as shown in Figure \ref{fig:eval_sens}. 
The fastest Optane SSD matches best (i.e., \pname{} achieves the most benefit) with Fast network, 
the slower Intel DC SSD matches with Medium network, and the slowest P-SSD matches with Slow network. 
%Though \pname{} brings less benefit under unmatched network and storage latencies, it is less a concern 
The reduced benefit for \mbox{\pname{}} under unmatched latencies is a potential limitation, but this is less 
of a concern, as modern data centers usually upgrade network and storage hardware together for best resource efficiency (e.g., using slow 
storage with fast RDMA network is impractical in the real world). Therefore, 
pairing the storage stack with the network stack fully unleashes the potential of \pname{}.

\subsubsection{Software-Isolated vSSDs vs. Hardware-Isolated vSSDs}

To examine \mbox{\pname{}} for software-isolated vSSDs, we run two software-isolated vSSDs on the same flash
channels (SW-Isolated). These vSSDs are isolated using token bucket rate limiting and both run YCSB with 50\% writes.  
%while the other vSSD runs a YCSB workload that is either read-only (SW-Isolated R-O) or write-only
%(SW-Isolated W-O). 
We compare SW-Isolated with a hardware-isolated vSSD (HW-Isolated) that has the full ownership of 
the flash channels. 

\mbox{\pname{}} reduces the P99.9 latency by 1.47$\times$ in comparison with VDC
for SW-Isolated vSSDs and by 1.51$\times$ for HW-Isolated vSSDs, 
as shown in Figure~\mbox{\ref{fig:eval_sw_hw_cdf}}. 
With hardware-isolated vSSDs, \mbox{\pname{}} brings marginally more benefit, since the hardware-isolated vSSD minimizes the 
interference from colocated workloads. 
%The hardware-isolation can provide marginally
%better coordinated I/O scheduling since there is no interference from a colocated workload. 
%We observe similar GC behavior and similar speedups from coordinated GC since
%the overall write ratio is the same for both settings.
%Different from hardware-isolated vSSDs, software-isolated vSSDs may have interference from the colocated vSSD. Therefore, with coordinated 
%GC, \mbox{\pname{}} can bring more performance benefits for software-isolated vSSDs than hardware-isolated vSSDs. 
%For instance, as for the 99.5th percentile read latency, \mbox{\pname{}} outperforms VDC by 3.4$\times$ with software-isolated vSSDs, and 
%by 1.5$\times$ with hardware-isolated vSSDs. 
%For example, at the 99.5-th percentile read tail latency, we see that SW-Isolated W-O 
%has $3.0\times$ worse performance than hardware isolation with VDC and only $1.7\times$ worse
%performance with \mbox{\pname{}}. 
Thus, \mbox{\pname{}} can improve the performance for both software-isolated and hardware-isolated vSSDs with the  
coordinated I/O scheduling and GC.

\vspace{-1ex}
\subsection{Benefits of Rack-Scale Wear Leveling}
\label{subsec:eval_wear}

\begin{figure}[t]
   \centering
   \includegraphics[width=0.9\linewidth]{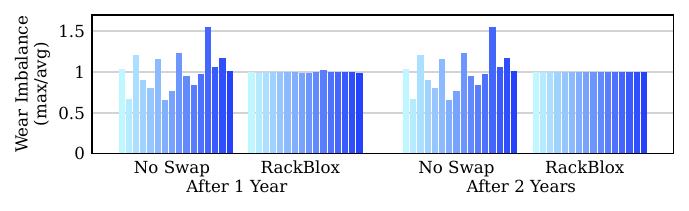}
   \vspace{-3ex}
   \caption{Benefits of \pname{} on each server's wear balance (different colors represent different SSDs in one server).}
   \label{fig:eval_wear_level_year}
   \vspace{-2ex}
\end{figure}
\begin{figure}[t]
   \centering
   \includegraphics[width=0.85\linewidth]{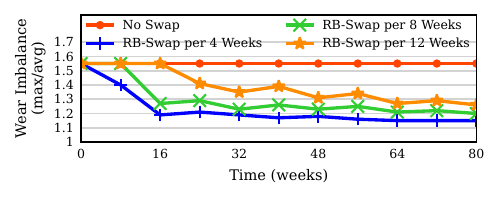}
   \vspace{-3ex}
	\caption{Benefits of \pname{} on rack-scale wear balance (the lower is better).}
   \label{fig:eval_wl_inter}
   \vspace{-2.0ex}
\end{figure}

To evaluate the rack-scale wear leveling of \pname{}, we simulate the effects of
running real workloads. 
We configure a rack with 32 servers, each server has 16 SSDs, and each SSD hosts 4 vSSDs. 
Each vSSD runs one workload (see Table~\ref{tab:real_workloads}), which may cause wear imbalance 
across different SSDs, since the workloads have diverse erase frequency. 
Each SSD is well balanced at the device level as it has its own device-level wear leveling.
Following the load balancing of modern storage infrastructures, we 
assign the vSSDs across servers using round robin~\cite{www-nginx}. We evaluate \pname{}'s
hierarchical wear leveling against modern storage infrastructure which does not swap across SSDs and 
servers (No Swap)~\cite{maneas:fast2020,maneas:fast2022}.

\noindent
\textbf{Local wear balancer.} Figure~\ref{fig:eval_wear_level_year} demonstrates that \pname{}'s
local wear balancer effectively maintains wear balance
across different SSDs. While No Swap
has significant wear imbalance, \pname{} can ensure near-optimal wear balance across the SSDs in
each server with periodic swapping.

\noindent
\textbf{Global wear balancer.} Local wear balancing suffers wear-imbalance at rack scale.
Figure~\ref{fig:eval_wl_inter} shows that \pname{}'s global wear balancer effectively
maintains rack-scale wear balance, despite reduced swapping frequency (e.g., 8
weeks). %\pname{} achieves wear balance over time.

\vspace{-0.5ex}

\vspace{-1ex}
\section{Related Work}
\label{sec:related}

\paragraph{Software-defined networking.} 
%Researchers have conducted research on SDN for decades. 
Recent studies investigated %almost every 
%aspect of the 
SDN systems, including networking abstraction, packet
processing and scheduling, QoS, SDN programming, performance, and fault
tolerance~\cite{pisces:sigcomm2016, pat:sigcomm2014, pifo:sigcomm2016, snap:sigcomm2016, fairqueue:nsdi2018}.
Programmable schedulers and frameworks have been proposed to allow developers to develop a variety of scheduling
algorithms~\cite{pifo:sigcomm2016, slack:algorithm}. 
%It is not surprising that most of the work centers around network techniques themselves. 
With these efforts, the community has produced a set of open-sourced
frameworks such as OpenFlow~\cite{www-openflow}, and the programming language P4~\cite{pat:sigcomm2014},
as well as the hardware devices like Intel Tofino~\cite{barefoot}. 
%which further motivates the development and deployment of SDN. 
Recent work~\cite{liu:asplos2017, netcache:sosp2017, netchain:nsdi2018} also
demonstrates that SDN can benefit distributed storage systems. %by implementing caching and
%data consensus protocols in programmable switches. 
However, none studied the codesign of SDN and SDS. %leaving much space for improvement. 
\pname{} makes an initial effort in this, and shows the benefits of the new software-defined 
rack-scale storage system. 
% holistic approach that combines SDN and SDS together to address the challenges in data centers.
%Moreover, researchers recently started to investigate in-network computing~\cite{sanvito:netcompute2018}
%for different use-cases, such as stream processing~\cite{jepsen:sosr2018} and query
%processing~\cite{lerner:cidr2019}, however, it is still not clear how should we architect
%in-network computing/accelerators with resource-constrained programmable switches, and how it will impact
%the large-scale data processing.

\noindent
\paragraph{Software-defined storage.} 
Researchers proposed techniques 
%for flash management 
like SDF~\cite{ouyang:asplos2014, wang:eurosys2014, lee:fast2016} and open-channel SSDs~\cite{bjorling:fast2017},
so upper-level system software can exploit the intrinsic properties of flash memory. %and to achieve improved performance isolation.
%The SDF decouples the flash translation layer and data access on flash chips, and treats them as the control plane and data plane.
As the cost of flash-based SSDs approaches that of HDDs and their performance has improved, 
SDF is a compelling solution for storage management in data centers~\cite{ouyang:asplos2014, huang:fast17}.
%There is no doubt that modern data centers have been moving forward to embrace software-defined storage. 
%Typical examples include the real software-defined networking devices like
%Barefoot Tofino~\cite{barefoot} and NetFPGA~\cite{www-netfpga}, and storage devices like open-channel
%SSDs~\cite{www-cnex-ssd}.
%Similar to SDN development, 
%Intensive studies have also been conducted on SDF techniques. 
However, no previous study focused on the integration of SDN and SDF.

\noindent
%\paragraph{Software-Defined Scheduling.}
\paragraph{Network/storage co-scheduling.}
To improve the end-to-end performance for data center applications, 
%the systems community has proposed solutions such as 
IOFlow~\cite{eno:sosp2013} and VDC~\cite{virtualdc:osdi2014} enforced policies
for I/O requests in centralized servers or hypervisors. However, they treated the SDN and SDF as black boxes
without considering the underlying hardware opportunities.
%To implement the concurrent data access in distributed storage, the community has developed both
%centralized~\cite{hastings1990distributed, yan2016leveraging, jula2008deadlock, huang2017top}
%and decentralized locking schemes~\cite{drtm, dslr, devulapalli2005distributed, narravula2007high}.
%However, they either suffer from scalability issue or lose the flexibility due to the implementation complexity.
%To achieve the fault tolerance, replication schemes~\cite{chain-replication, nopaxos} have been widely used
%in data centers. They have two replication protocols: primary-backup~\cite{primary-backup, chain-replication,
%fawn, flexkv, hyperdex} and quorum-based~\cite{paxos98, paxos01, raft}. Unfortunately, none of them can
%achieve both linear-scalability and strong data consistency.
Recently, researchers
leveraged programmable switches to fulfill system functions like data
caching~\cite{netcache:sosp2017}, consistency protocols~\cite{netchain:nsdi2018}, and task scheduling~\cite{racksched:osdi2020}, 
showing that it is feasible to integrate system functions into programmable switches. 
%However, these solutions were implemented
%for specific applications and distributed protocols.
%integration of SDF into SDN, investigate its challenges on system
%design and implementation, and demonstrate its benefits brought to the storage management and
%end-to-end performance.
%As discussed, the \textbf{\textit{industry community}} has produced programmable
%switches such as Barefoot Tofino and programmable flash-based SSDs such as Open Channel SSD~\cite{www-cnex-ssd,
%jian:fast2017, bjorling:fast2017} to accelerate the deployment of SDN and SDF in data centers~\cite{sdn:acm2016}.
We integrate storage functions into SDN and show the benefits of this design.  
%the innovation of 
%a new software-defined infrastructure with network/storage
%co-design. %We will also demonstrate its benefits to a diversity of applications while preserving the programmability.

\vspace{-1ex}

\vspace{-1ex}
\section{Conclusion}
\label{sec:conclusion}
We present \pname{}, a new rack-scale storage system by co-designing the software-defined 
networking and storage stack. \pname{} integrates essential storage functions
into the programmable switch, and enables the state sharing between the network and storage stack. 
With coordinated I/O scheduling, GC, and rack-scale wear leveling, 
\pname{} achieves improved end-to-end storage performance, while ensuring near-ideal lifetime 
for SSDs in a rack.  

\vspace{-1ex}
\begin{acks}
We thank the anonymous reviewers and our shepherd Simon Peter for their insightful comments and feedback. 
We thank Yiqi Liu for proofreading an early version of this paper. %This work was partially supported by NSF grant CCF-1919044, 
%We thank Yiqi Liu for his help at the early stages of this work. 
This work was partially supported by NSF grant CCF-1919044, 
CCF-2107470, and the Hybrid Cloud and AI program at the IBM-Illinois Discovery Accelerator Institute. 
\end{acks}

%\newpage
\balance
\small
\bibliographystyle{plain} 
\bibliography{iscaref,references,network}

\end{document}